\documentclass[journal]{IEEEtran}
\usepackage{cite}
\usepackage{mathtools}
\usepackage{amsthm,amsmath,amssymb}
\usepackage{amssymb}
\usepackage{mathrsfs}
\usepackage{graphicx}
\usepackage{epstopdf}
\usepackage{float} 
\usepackage{multirow}
\usepackage{booktabs}
\usepackage{url}
\usepackage{bm}
\usepackage{array} 

\usepackage[dvipsnames]{xcolor}
\usepackage{blkarray}
\graphicspath{{./figure/}}
\DeclareGraphicsExtensions{.pdf,.jpeg,.png,.jpg}

\usepackage{colortbl}
\usepackage[ruled,lined,linesnumbered]{algorithm2e}
\usepackage{graphicx}
\usepackage[caption=false,font=footnotesize]{subfig}

\usepackage[hidelinks]{hyperref}
\usepackage{booktabs}
\usepackage{tikz}
\usepackage{tabularx}

\ifCLASSINFOpdf
\else
\fi

\graphicspath{{figures/}}
\usepackage{amsmath} 
\usepackage{ragged2e}

\begin{document}

\title{Exploiting the Flexibility of LLM Inference for Frequency Regulation Service}

\author{Yimeng Sun,~\IEEEmembership{Student Member,~IEEE,}
       and Zhaohao Ding,~\IEEEmembership{Senior Member,~IEEE}\\}
        
\IEEEaftertitletext{\vspace{-0.5cm}}
\maketitle

\noindent
\begin{minipage}{\columnwidth}
\footnotesize
\textcopyright{} 2026 IEEE. Personal use of this material is permitted.  Permission from IEEE must be obtained for all other uses, in any current or future media, including reprinting/republishing this material for advertising or promotional purposes, creating new collective works, for resale or redistribution to servers or lists, or reuse of any copyrighted component of this work in other works.
\end{minipage}
\par\vspace{0.5\baselineskip}

\begin{abstract}
While the widespread deployment of Large Language Models (LLMs) has significantly driven up power demands in computing clusters, the unique execution characteristics of LLM inference offer a novel opportunity to utilize these clusters as controllable loads. In this paper, we investigate the potential of LLM inference cluster to provide frequency regulation services through inference request scheduling. First, we model the problem as a hierarchical Markov decision process that includes request-to-replica dispatching and token-level resource allocation. Then, we introduce a reward redistribution network to address the temporal asynchrony between millisecond-level scheduling decisions and second-level frequency regulation signals. Furthermore, we develop an optimization framework based on a mean-field approximation to enable coordination among a large number of replicas under high request concurrency. Simulations using real-world data demonstrate that the proposed approach accurately tracks frequency regulation signals without compromising service-level objectives.
\end{abstract}


\IEEEpeerreviewmaketitle

\section*{Nomenclature}
\renewcommand{\arraystretch}{1.08}

\noindent\textit{\textbf{Indices and Sets}}

\vspace{3mm}
\begin{tabularx}{\columnwidth}{@{}lX@{}}
$i$ & Index of LLM replicas. \\
$j$ & Index of inference requests. \\
$t$ & Index of fine-grained scheduling decision steps. \\
$k$ & Index of regulation intervals. \\
$\mathcal{T}_k$ & Set of decision steps belonging to regulation interval $k$. \\
$B_{t,d}$ & Reward segment starting at step $t$ and containing $d$ decision steps. \\
$\mathcal{B}$ & Non-overlapping partition of an episode trajectory into reward segments. \\
\end{tabularx}

\vspace{1mm}
\noindent\textit{\textbf{Variables}}

\vspace{1mm}
\begin{tabularx}{\columnwidth}{@{}lX@{}}
$s_t^u$, $a_t^u$ & Upper-layer state and request-dispatching action at step $t$. \\
$s_{i,t}^l$, $a_{i,t}^l$ & State and token-allocation action of replica $i$ at step $t$. \\
$\boldsymbol{s}_t^l$, $\boldsymbol{a}_t^l$ & Joint state and joint action of all lower-layer replicas. \\
$p_{i,t}$ & Power consumption of replica $i$ at step $t$. \\
$\bar{P}_k$ & Average aggregate power consumption during regulation interval $k$. \\
$P_k^{\mathrm{tar}}$ & Target regulation power during interval $k$. \\
$\bar{a}_{-i,t}^l$ & Mean action of all replicas other than replica $i$. \\
\end{tabularx}

\begin{tabularx}{\columnwidth}{@{}lX@{}}
$r_t$, $\hat{r}_t$ & Original and redistributed rewards at step $t$, respectively. \\
$G_t$ & Discounted return from decision step $t$. \\
$R_{t,d}^{\gamma}$ & Discounted original reward over segment $B_{t,d}$. \\
$A_{i,t}$ & Advantage estimate of lower-layer replica $i$. \\
\end{tabularx}

\vspace{1mm}
\noindent\textit{\textbf{Parameters}}

\vspace{2mm}
\begin{tabularx}{\columnwidth}{@{}lX@{}}
$N$ & Number of LLM replicas. \\
$T$ & Number of decision steps in an episode. \\
$K$ & Number of regulation intervals in an episode. \\
$D$ & Number of decision steps in each regulation interval, with $T=KD$. \\
$M$ & Maximum number of local requests considered by each replica. \\
$\kappa_i$ & Maximum number of tokens that replica $i$ can process in one decision step. \\
$\gamma$ & Discount factor. \\
$\lambda_1$, $\lambda_2$ & Penalty coefficients for TTFT and E2E violations. \\
$\eta$ & Clipping parameter of PPO. \\
$\theta^u$ & Parameter vector of the upper-layer DQN. \\
$\theta_i^l$, $\psi_i^l$ & Actor and critic parameters of lower-layer replica $i$. \\
$\phi_R$ & Parameter vector of the reward redistribution network. \\
$\alpha_{u,q}$, $\alpha_{l,q}$ & Upper- and lower-layer learning rates at iteration $q$. \\
\end{tabularx}

\vspace{1mm}
\noindent\textit{\textbf{Functions, Policies, and Models}}

\vspace{2mm}
\begin{tabularx}{\columnwidth}{@{}lX@{}}
$\Gamma$ & Hierarchical MDP describing the two-stage inference scheduling process. \\
$\mathcal{P}$ & State-transition function. \\
$\pi$ & Joint hierarchical scheduling policy. \\
$\pi_i^l$ & Local token-allocation policy of replica $i$. \\
$Q^u$ & Upper-layer action-value function used for request dispatching. \\
$Q_i^l$ & Mean-field action-value function of replica $i$. \\
$f_R$ & Reward redistribution network. \\
$\mathcal{L}_{\mathrm{redis}}$ & Reward redistribution loss. \\
$\mathcal{L}_i^{\mathrm{critic}}$ & Critic loss of replica $i$. \\
$\mathcal{L}_i^{\mathrm{actor}}$ & Clipped PPO actor loss of replica $i$. \\
$\mathcal{M}$, $\hat{\mathcal{M}}$ & Original and reward-redistributed MDPs, respectively. \\
\end{tabularx}

\vspace{1mm}
\noindent\textit{\textbf{Superscripts and Subscripts}}

\vspace{2mm}
\begin{tabularx}{\columnwidth}{@{}lX@{}}
$u$, $l$ & Upper and lower scheduling layers, respectively. \\
$p$, $d$ & Prefill and decode phases, respectively. \\
$-i$ & Aggregate quantity associated with all replicas except replica $i$. \\
$\mathrm{NE}$ & Stationary mean-field Nash equilibrium. \\
\end{tabularx}

\section{Introduction}

\IEEEPARstart{T}{he} rapid development of artificial intelligence (AI), led by large language models (LLMs), is driving a sharp increase in the energy consumption of computing clusters. Global AI energy consumption is projected to double from 415 TWh in 2024 to approximately 945 TWh by 2030~\cite{iea2024energyai}. Concurrently, the dominant energy burden is shifting from model training to inference, with the latter expected to account for over 60\% of total AI energy consumption by 2026~\cite{huawei2024intelligent}. This paradigm shift presents an opportunity to exploit the unique execution characteristics of LLM inference and allow its computing cluster to function as a controllable load. Specifically, the LLM inference process consists of two sequential phases, namely prefill and decode, both of which exhibit inherent temporal flexibility~\cite{gao2025apt}. In many application scenarios, such as speech-to-text transcription and image understanding, latency of several seconds can be tolerated without degrading the user-perceived quality of experience. In more complex scenarios, such as chain-of-thought reasoning and AI agent, even longer latencies are acceptable~\cite{openai2024deepresearch,guo2025deepseek}. Consequently, the LLM inference cluster could serve as a promising flexible demand-side resource, capable of adjusting its power load and providing frequency regulation services to the power grid.

In practice, power regulation can be achieved through both hardware- and software-level control. Hardware-level methods, such as dynamic voltage and frequency scaling (DVFS) and GPU power capping, directly adjust device power to provide regulation response~\cite{li2013data,zhou2026grid,jahanshahi2022powermorph}. Software-level methods instead reshape the aggregate power profile by scheduling when and how workloads are executed while satisfying service requirements. These approaches exploit complementary sources of flexibility. This paper focuses on software-level control and harnesses the temporal flexibility of LLM inference for frequency regulation without directly modifying GPU power limits.

Although limited efforts have been dedicated to exploiting LLM inference flexibility for frequency regulation, extensive research has investigated the potential of utilizing general computing workloads for such services. Chen~\textit{et al.}~\cite{chen2019energyqare} propose rule-based control frameworks that combine finite-state scheduling mechanisms with state feedback to balance quality of service and regulation tracking performance. Zhang~\textit{et al.}~\cite{zhang2021hpc} and Chen~\textit{et al.}~\cite{chen2013dynamic} explore job scheduling and server-level power capping strategies in high-performance computing settings, where generalized processor sharing is employed to allocate resources and follow secondary frequency signals. Zhang~\textit{et al.}~\cite{zhang2024data} and Pahlevan~\textit{et al.}~\cite{pahlevan2020ecogreen} examine virtual machine consolidation and inter-data center workload migration with the goal of minimizing electricity costs while enhancing coordination capabilities. Clark~\textit{et al.}~\cite{clark2024learning} incorporate learning-based control models to improve adaptability and scalability in dynamic regulation environments. Jahanshahi \textit{et al.} \cite{jahanshahi2022powermorph} introduce complementary workloads as a flexible control dimension to improve regulation capacity while maintaining the performance of latency-sensitive services. Li \textit{et al.} \cite{li2014integrated} propose a hierarchical control architecture that coordinates servers and electric vehicle fleets to jointly participate in the regulation and energy markets. However, these existing workload-level scheduling methods generally operate at the granularity of jobs, servers, or virtual machines and do not capture the fine-grained execution characteristics of LLM inference. Unlike general computing workloads, effectively exploiting the temporal flexibility of LLM inference requires token-level scheduling across its prefill and decode phases. Therefore, conventional coarse-grained scheduling methods are not directly suited to LLM inference scheduling for frequency regulation.

More recently, fine-grained execution and resource management for LLM inference have been explored to improve service performance and energy efficiency. At the execution level, token scheduling and adaptive batching coordinate different inference stages and balance latency and energy consumption~\cite{wilkins2024online,argerich2024measuring}. At the system level, workloads are coordinated across cloud-edge and heterogeneous platforms to improve resource utilization and reduce energy costs~\cite{yang2024perllm,wilkins2024offline}. Other studies further reduce energy consumption through model and scheduling optimization, adaptive model parallelism and GPU frequency, or renewable-aware execution~\cite{hisaharo2024optimizing,stojkovic2025dynamollm,khoshsirat2024decentralized}. Collectively, these studies demonstrate the potential of fine-grained inference management to reshape resource utilization and energy consumption. Nevertheless, their objectives remain centered on service performance and energy efficiency. They neither treat LLM inference clusters as controllable grid resources nor schedule their aggregate power demand to track frequency regulation signals.

To empower the LLM inference cluster with the capability for frequency regulation, two main challenges need to be addressed. The first challenge arises from \textbf{\textit{asynchrony between decision making and feedback}}, introduced by the intrinsic mismatch in timescales between LLM inference scheduling and regulation signals. While LLM inference operates at millisecond-level intervals, regulation signals are typically updated every two to four seconds. This temporal asynchrony results in coupled and delayed feedback, where high-frequency scheduling decisions interact over time before any impact on regulation-tracking performance can be assessed. Existing inference scheduling methods, which often treat decisions as isolated and instantaneous events, fail to capture these asynchronous interactions and multi-timescale dynamics. Therefore, they struggle to effectively utilize the fine-grained flexibility of LLM inference for accurate frequency regulation. The second challenge lies in \textbf{\textit{large-scale coordination}} during the LLM inference process. Modern LLM inference clusters usually handle tens of thousands of requests per second on large GPU servers \cite{supermicro2024xaicolossus}. At this scale, optimizing scheduling decisions induces a prohibitive computational burden. Current approaches often struggle to manage the complexity, which limits their capacity to efficiently coordinate massive inference requests with regulation signals without sacrificing control accuracy.

To tackle the above challenges, we propose a hierarchical reinforcement learning framework that enables LLM inference cluster to provide frequency regulation services via inference request scheduling. It integrates a reward redistribution mechanism to manage delayed and coupled rewards. In addition, a mean-field approximation–based structure is employed to enable scalable coordination in large-scale scenarios. The main contributions of this paper are summarized as follows.

\textit{1)} We establish a hierarchical Markov decision process (MDP) framework to model fine-grained scheduling of LLM inference requests for frequency regulation. The upper layer handles request dispatching, while the lower layer performs token-level resource allocation control.

\textit{2)} We propose a dynamic reward redistribution mechanism based on a cross-attention structure to reassign delayed and coupled rewards to corresponding decision steps. The mechanism improves scheduling stability and regulation performance, and we prove that the reconstructed MDP remains approximately equivalent to the original formulation.

\textit{3)} We develop a large-scale reinforcement learning algorithm based on mean-field approximation for LLM inference scheduling, which enables scalable multi-agent coordination and providing a theoretical proof of convergence.

The remainder of this paper is organized as follows. Section II models the frequency regulation problem by leveraging LLM inference request scheduling. Section III presents a multi agent reinforcement learning (MARL)-based solution for large-scale request scheduling with delayed feedback. Section IV presents the results of our numerical analyzes and Section V provides the conclusions.

\section{Problem Formulation}

\subsection{Problem Description}

\begin{figure}[bht]
\vspace{-3mm}
\centering
\includegraphics[width=7cm]{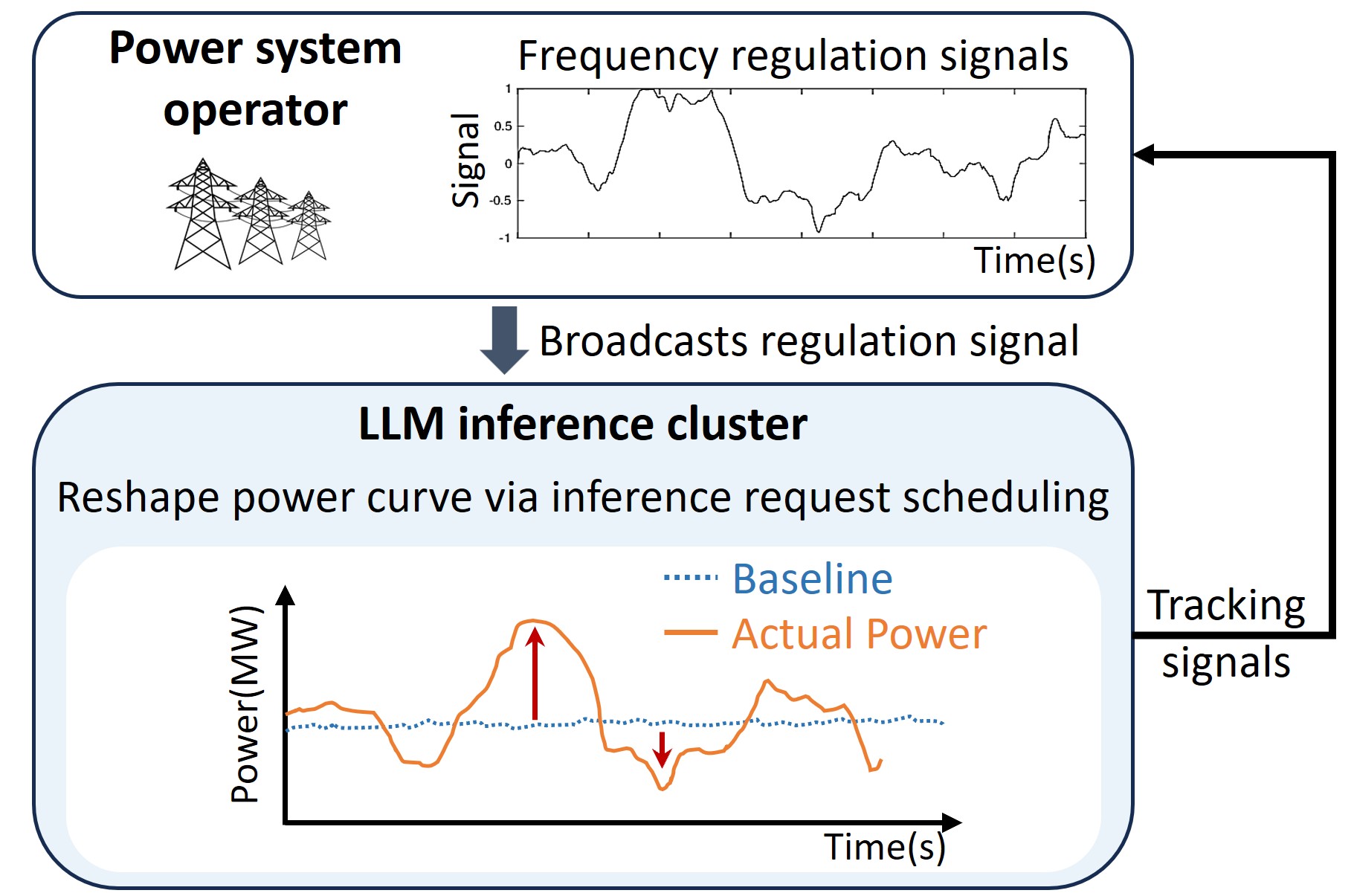}
\vspace{-3mm}
\caption{Schematic of LLM inference cluster providing frequency regulation.}
\label{fig:framework}
\end{figure}

We investigate the use of LLM inference flexibility to provide AGC-based secondary frequency regulation. Specifically, this study focuses on the real-time response stage of the ancillary service market, where the system operator periodically broadcasts a time-varying target power signal after regulation capacity has been assigned. In response, the LLM inference cluster adjusts its power consumption through a two-stage scheduling process to track the signal while maintaining inference SLOs, as shown in Fig.~\ref{fig:framework}.

\begin{figure}[t]
\vspace{-3mm}
\centering
\includegraphics[width=7cm]{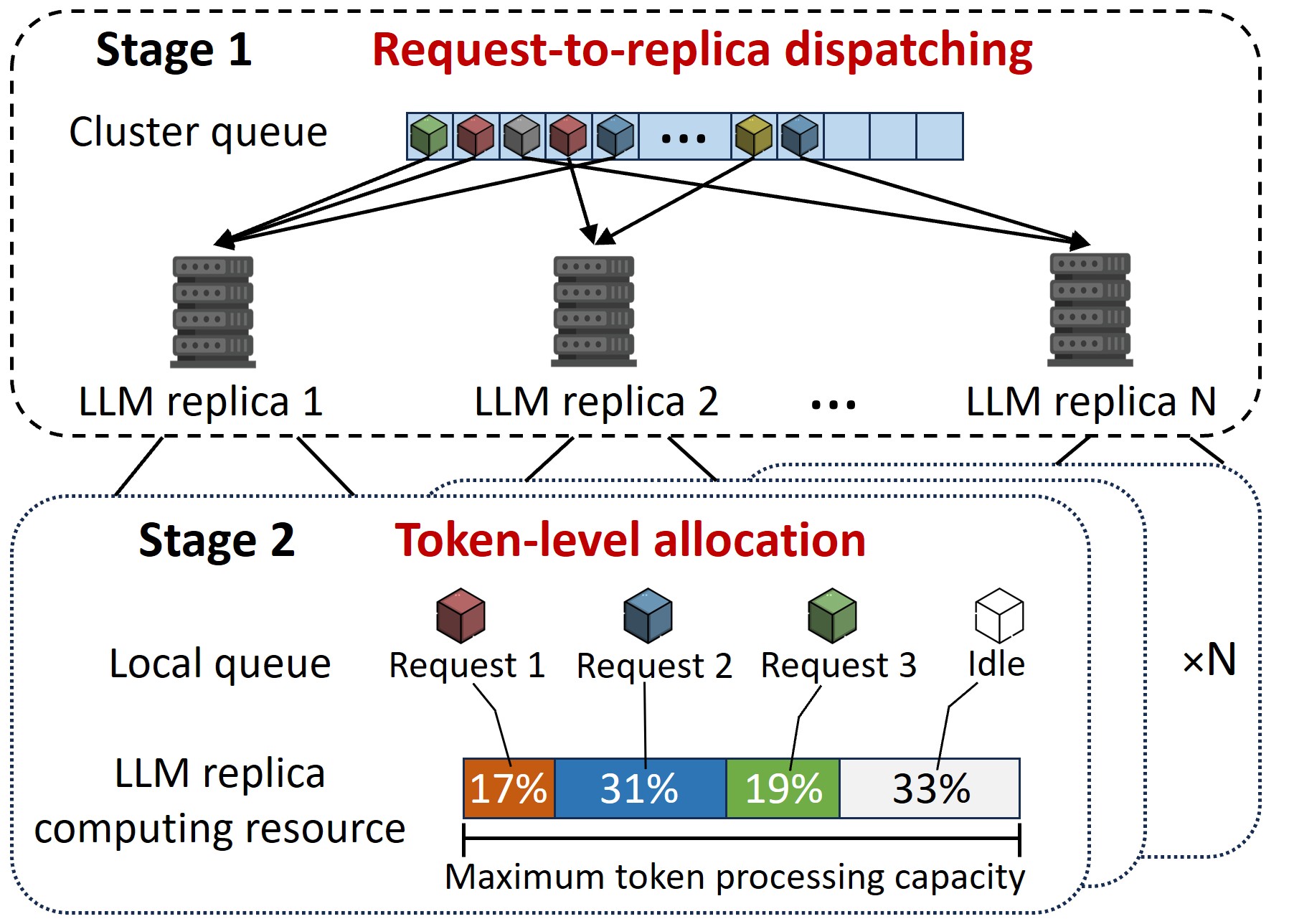}
\caption{Two-stage scheduling process for LLM inference requests.}
\label{framework2}
\end{figure}

Specifically, we model the two-stage scheduling process as shown in Fig.~\ref{framework2}. The LLM inference cluster consists of multiple LLM replicas, where each replica is a copy of the same model running in parallel on GPU servers to serve inference requests~\cite{googleVLLMDeployment,nvidiaNIMScale}. In the first stage, massive inference requests arrive in the cluster queue. The cluster scheduler then dispatches these requests to specific replicas based on the current utilization states of the replicas~\cite{srivatsa2025preble}. In the second stage, dispatched requests enter the local queue of each replica. The replica scheduler then performs fine-grained, token-level resource allocation~\cite{yu2022orca,kwon2023vllm,agrawal2024sarathi}. As discussed earlier, the execution of inference requests comprises the prefill and decode phases, both of which can tolerate a certain amount of delay~\cite{zhong2024distserve}. These delays are constrained by the time-to-first-token (TTFT) and end-to-end (E2E) latency requirements, respectively~\cite{gao2025apt}. These scheduling decisions alter the batch composition and token processing rate of each replica, which in turn produce time-varying GPU execution intensity and power consumption. Notably, even under identical resource allocations, the prefill and decode phases exhibit different power characteristics due to their distinct computational patterns~\cite{patel2024characterizing}. To capture such structural heterogeneity, we explicitly model the two phases separately in the subsequent formulation. Then, the aggregate power consumption of all replicas constitutes the total power load of the inference cluster. By coordinating millisecond-level scheduling decisions across the two stages, the power load of the inference cluster can be dynamically regulated to provide frequency regulation services. Specifically, the average power over each regulation interval is regarded as the effective response to the target regulation signal. Under this framework, our objective is to minimize the tracking error of the frequency regulation signal while satisfying the SLOs of LLM inference.

\subsection{Hierarchical MDP Formulation}

We formulate the two-stage scheduling process as a hierarchical MDP, defined by the tuple 
$\Gamma = \langle \mathcal{S}^u, \mathcal{S}^l, \mathcal{A}^u, \mathcal{A}^l, \mathcal{P}, \mathcal{R}, \gamma \rangle$. The cluster scheduler acts as the upper-layer agent, while $N$ replica schedulers, indexed by $i \in \{1, \dots, N\}$, serve as the lower-layer agents. We use $t\in\{1,\ldots,T\}$ to index the fine-grained scheduling decision steps and $k\in\{1,\ldots,K\}$ to index the regulation 
intervals. Each regulation interval contains $D$ consecutive decision steps, such that $T=KD$. For a given decision step $t$, its corresponding regulation interval is indexed by $k=\lceil t/D\rceil$, and the set of decision steps within interval $k$ is denoted by $\mathcal{T}_k=\{(k-1)D+1,\ldots,kD\}$. The target regulation power for interval $k$ is denoted by $P_k^{\mathrm{tar}}$ and remains unchanged for all $t\in\mathcal{T}_k$.

\textbf{\textit{Upper-layer State}}: The upper-layer state $s^u_t$ consists of two parts: the status of all replicas $s^{rep}_t$ and the features of the cluster queue $s^{req}_t$. For each replica $i$, the status $s^{rep}_t$ includes the number of pending prefill tokens $\rho_i^p$, pending decode tokens $\rho_i^d$, and the counts of TTFT and E2E violations, denoted by $\omega_{i,t}^{\mathrm{TTFT}}$ and $\omega_{i,t}^{\mathrm{E2E}}$, respectively. $s^{req}_t$ contains an embedding vector $e^{req}_t$ that summarizes all requests currently waiting in the queue, as well as the token demands of the current request $j$, represented by $\rho_j^p$ and $\rho_j^d$.

\textbf{\textit{Upper-layer Action}}: The upper-layer action $a^u_t$ determines which replica is selected to handle the request $j$. It is chosen from a discrete action space $\{1, 2, \dots, N\}$, where each element corresponds to the index of a specific LLM replica.

\textbf{\textit{Lower-layer State}}: After the upper-layer action is executed, the selected request $j$ is added to the local queue of the assigned replica $i$. The state of replica $i$ at time $t$, denoted by $s^l_{i,t}$, is composed of two parts: the local information $s^{\text{loc}}_{i,t}$ and the cooperative statistics $s^{\text{coop}}_{i,t}$. The local information includes the token demands of the current request ($d^p_j$, $d^d_j$), the numbers of pending prefill and decode tokens ($\rho^p_{i,t}$, $\rho^d_{i,t}$), the TTFT and E2E violation counts ($\omega_{i,t}^{\mathrm{TTFT}}$, $\omega_{i,t}^{\mathrm{E2E}}$), the current power consumption $p_{i,t}$, the target regulation powers for the current and $L$ preceding intervals $(P_k^{\mathrm{tar}},P_{k-1}^{\mathrm{tar}},\ldots,
P_{k-L}^{\mathrm{tar}})$, and the history of tracking errors over $H$ previous time steps $(e_{t-1}, \dots, e_{t-H})$. The cooperative statistics describe the average operational state of the other replicas, including the average power $\bar{p}_{-i,t}$, average pending tokens ($\bar{\rho}^p_{-i,t}$, $\bar{\rho}^d_{-i,t}$), and average violation counts ($\bar{\omega}_{-i,t}^{\mathrm{TTFT}}$ , $\bar{\omega}_{-i,t}^{\mathrm{E2E}}$).

\textbf{\textit{Lower-layer Action}}: Each replica scheduler performs token-level resource allocation for requests in its local queue. At each time step, replica $i$ allocates a limited token budget among $M$ local requests, with an additional null action representing unassigned tokens. The action $a^l_{i,t} \in \mathbb{R}^{M+1}$ denotes the token allocation vector and satisfies the capacity constraint:
\begin{equation}
\sum_{m=1}^{M+1} a^l_{i,t}[m] = \kappa_i, \quad a^l_{i,t}[m] \ge 0,
\end{equation}
where $\kappa_i$ is the maximum number of tokens that replica $i$ can process in one step.

\textbf{\textit{Transition Function}}: The transition function $\mathcal{P}(s^u_t, \{s^l_{i,t}\}, a^u_t, \{a^l_{i,t}\}) \rightarrow (s^u_{t+1}, \{s^l_{i,t+1}\})$ describes the state updates induced by both layers' action. Once the upper layer dispatches request $j$ to replica $i$, it is added to the local queue. The lower layer then allocates token-level resources and deterministically updates token progress, replica states, and violation metrics.

\emph{\textbf{Reward and $\gamma$}}:  
The reward at time step $t$ depends jointly on the upper-layer dispatching and lower-layer allocation decisions. It measures overall system performance in frequency regulation and inference responsiveness.  
The total reward is defined as:
\begin{equation}
r_t
=
w_{\mathrm{fre}}r_t^{\mathrm{freq}}
+
w_{\mathrm{SLO}}r_t^{\mathrm{SLO}}
+
w_{\mathrm{com}}r_t^{\mathrm{comp}},
\end{equation}
where the three components are as follows.

\begin{itemize}
\item 
\textbf{Regulation signal tracking error} ($r_t^{\mathrm{freq}}$):

Let $P^{\mathrm{base}}$ denote the nominal cluster power without regulation actions, which serves as the reference operating point in the simulation. Let $P^{\mathrm{rated}}$ denote the aggregate rated power of the GPUs in the cluster, and let $R_{\mathrm{fre}}$ denote the regulation capacity ratio. Given the normalized regulation signal $u_k\in[-1,1]$, the target power during interval $k$ is defined as
\(
P_k^{\mathrm{tar}}
=
P^{\mathrm{base}}
+
u_kR_{\mathrm{fre}}\times P^{\mathrm{rated}}.
\)

This term evaluates how closely the aggregate power consumption of all replicas tracks the target regulation power over each regulation interval. For each interval $k$, the average aggregate power and the corresponding tracking reward are defined as
\begin{equation}
\begin{aligned}
\bar{P}_k
&=
\frac{1}{D}
\sum_{t\in\mathcal{T}_k}
\sum_{i=1}^{N}p_{i,t},\\
r_t^{\mathrm{freq}}
&=
\begin{cases}
-\displaystyle\left|
\frac{\bar{P}_k-P_k^{\mathrm{tar}}}
{P_k^{\mathrm{tar}}}
\right|, & t=kD,\\[2mm]
0, & \text{otherwise}.
\end{cases}
\end{aligned}
\end{equation}
Here, $\bar{P}_k$ denotes the actual average power response of the inference cluster during interval $k$.

It is worth noting that $r_t^{\mathrm{freq}}$ is an interval-level delayed reward that is issued only at the end of interval $k$. It aggregates the cumulative effects of the fine-grained scheduling decisions within $\mathcal{T}_k$. Consequently, the contribution of each individual decision within the interval to the final tracking accuracy is not directly observable.

\item \textbf{SLO violation penalty} ($r_t^{\mathrm{SLO}}$):  
This term penalizes latency violations in both the prefill and decode phases:
\begin{equation}
r_t^{\mathrm{SLO}}
= -\lambda_1\sum_{i=1}^{N}\omega_{i,t}^{\mathrm{TTFT}}-\lambda_2\sum_{i=1}^{N}\omega_{i,t}^{\mathrm{E2E}}.
\end{equation}
where the coefficients $\lambda_1$ and $\lambda_2$ determine the relative penalty weights for TTFT and E2E latency violations.

\item 
\textbf{Request-completion reward} ($r_t^{\mathrm{comp}}$):
A small positive reward is issued for each request completed before its deadline:
\begin{equation}
r_t^{\mathrm{comp}}
=
\sum_{i=1}^{N}
\sum_{j\in\mathcal{J}_{i,t}^{\mathrm{comp}}}
r_j^{\mathrm{comp}},
\end{equation}
where $\mathcal{J}_{i,t}^{\mathrm{comp}}$ denotes the set of requests completed by replica $i$ at time $t$ without violating their deadlines, and $r_j^{\mathrm{comp}}>0$ is the completion reward for request $j$. This auxiliary reward encourages early, deadline-compliant completion.
\end{itemize}

The discounted return from decision step $t$ is defined as \(G_t=\sum_{\tau=t}^{T}\gamma^{\tau-t}r_\tau. \) The objective is to maximize the expected return from the beginning of an episode \(J(\pi)=\mathbb{E}_{\pi}[G_1], \) where $\pi$ denotes the joint hierarchical policy and $\gamma\in(0,1]$ is the discount factor.

\section{Solution Method}
The aforementioned MDP formulation provides a fine-grained model for the cluster in providing frequency regulation services. 
However, solving the proposed MDP still presents two major challenges, as discussed in Section I. The first challenge lies in the temporal asynchrony between decision-making and feedback. The second challenge is the exponential growth of the problem scale as the number of replicas increases. To tackle these challenges, we first design a reward redistribution network that reformulates the MDP by assigning coupled and delayed rewards to corresponding decision steps. Subsequently, we propose a large-scale optimization method based on mean-field approximation to achieve efficient coordination among replicas.

\subsection{Reward Redistribution under Asynchronous Decision-Making and Feedback}
We introduce a reward redistribution network to reassign interval-level regulation feedback to individual decision steps along the scheduling trajectory. By learning a mapping from the sequence of states and actions to step-wise reward contributions, the network decomposes delayed and aggregated regulation rewards into temporally localized signals. These signals provide denser and more informative feedback for policy optimization.

First, we define a reward segment as a contiguous sub-trajectory of the full trajectory, denoted by
\begin{equation}
\begin{aligned}
B_{t,d}
=
\left\{
\left(
s_\tau^u,
a_\tau^u,
\{s_{i,\tau}^l,a_{i,\tau}^l\}_{i=1}^{N}
\right)
\right\}_{\tau=t}^{t+d-1},\\
1\leq t\leq t+d-1\leq T,\
d\in\mathbb{Z}_{>0}.
\end{aligned}
\end{equation}
with the corresponding segment reward defined as $R_{t,d}^{\gamma}
=
\sum_{\tau=t}^{t+d-1}
\gamma^{\tau-t}r_\tau.$ where $r_\tau$ denotes the per-step reward at time $\tau$.

We then establish the condition under which the redistributed rewards preserve the optimality of the original MDP.

\textit{\textbf{Lemma 1 (MDP Equivalence under Reward Redistribution):}}
\textit{
Let \(\mathcal{M}=\langle\mathcal{S},\mathcal{A},r,\mathcal{P},\gamma\rangle\)
and
\(\hat{\mathcal{M}}=\langle\mathcal{S},\mathcal{A},\hat r,\mathcal{P},\gamma\rangle\)
denote the original and redistributed MDPs, respectively, with the same
initial state \(s_0\), where \(\gamma\in(0,1]\).
Suppose that an episode trajectory is partitioned into non-overlapping
reward segments. If every segment \(B_{t,d}\), with \(d>0\), satisfies
\(
\sum_{\tau=t}^{t+d-1}\gamma^{\tau-t}\hat r_\tau
=
R_{t,d}^{\gamma},
\)
then, for any policy \(\pi\),
\(
V^\pi_{\mathcal{M}}(s_0)
=
V^\pi_{\hat{\mathcal{M}}}(s_0).
\)
Consequently, the two MDPs share the same set of optimal policies.
}

\begin{proof}
See Appendix~A for the complete proof.
\end{proof}

\textit{Lemma~1 }shows that, if the redistributed rewards satisfy the segment-level return-conservation condition, the resulting MDP preserves the same optimal policy as the original process. Notably, this condition constrains only the discounted return over each segment rather than prescribing a unique reward for every individual decision step. To satisfy this constraint in practice, we employ an attention-based network that learns to generate stepwise redistributed rewards $\hat{r}_\tau$, ensuring that their discounted cumulative return approximates the total segment reward $R_{t,d}^{\gamma}$. 

\begin{figure}[t]
    \centering
    \includegraphics[width=8.4cm]{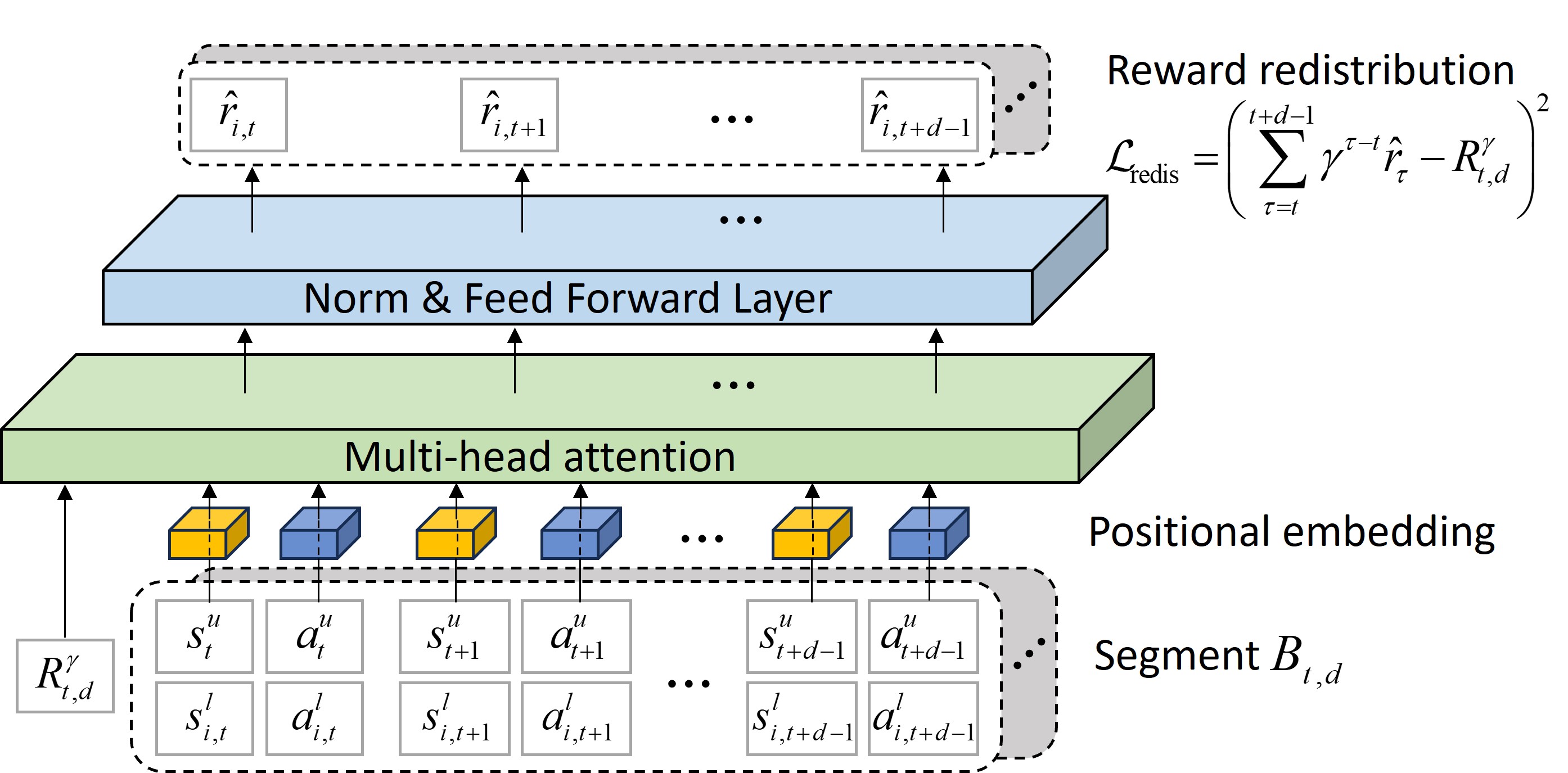}
    \caption{Framework of the proposed reward redistribution network.}
    \label{reward redistribution network}
    \vspace{-3mm}
\end{figure}

As shown in Fig.~\ref{reward redistribution network}, the network takes the state-action sequence within a segment as input and outputs the redistributed rewards for each step. Specifically, each state--action pair is encoded together with its positional information, while the cumulative segment reward serves as a global conditioning signal in the cross-attention module to capture contextual relationships with individual state--action representations. The training objective is to minimize the redistribution error:
\begin{equation} \mathcal{L}_{\mathrm{redis}} = \left( \sum_{\tau=t}^{t+d-1} \gamma^{\tau-t}\hat{r}_\tau - R_{t,d}^{\gamma} \right)^2.\end{equation} Here, Eq.~(7) enforces discounted return conservation at the segment level, while the cross-attention network parameterizes the context-dependent stepwise redistribution. After the redistribution network generates
\(\{\hat r_t,\ldots,\hat r_{t+d-1}\}\), the original reward field
\(r_\tau\) in each corresponding transition is replaced by
\(\hat r_\tau\), while its state, action, and next-state fields remain
unchanged. The reward-replaced transitions form a redistributed training
buffer \(\mathcal{D}'\). Both the upper-layer and lower-layer policy are trained using the redistributed transitions in \(\mathcal{D}'\).

Since the segment-sum constraint is approximated via learning, small deviations are inevitable. We quantify the impact of this approximation error on the final policy performance below.

\textbf{\textit{Theorem~1 (Performance Bound with Learned Redistribution):}}
\textit{Consider a trajectory partitioned into \(N_{\mathrm{seg}}\) non-overlapping reward segments
\(\mathcal{B}=\{B_{t_b,d_b}\}_{b=1}^{N_{\mathrm{seg}}}\), where \(d_b\) denotes the number of decision steps in segment \(b\). Suppose that the reward redistribution network is trained for \(n\) iterations using a first-order optimizer, and the empirical redistribution loss decreases at an
assumed sublinear rate of \(O(n^{-1/2})\). Then, for any policy \(\pi\), the value-function gap between the original and redistributed MDPs is bounded by}
\begin{equation}
\left|
V^\pi_{\mathrm{orig}}(s)
-
V^\pi_{\mathrm{redis}}(s)
\right|
\leq
C N_{\mathrm{seg}}n^{-1/4},
\end{equation}
\textit{where \(C\) is a constant proportional to the initial network error \(\sqrt{\mathcal{L}_{\mathrm{redis}}^{(0)}}\).}

\begin{proof}
See Appendix~B for the complete proof.
\end{proof}

\textit{Theorem~1 }shows that the value gap decreases as training proceeds and becomes negligible with sufficient iterations. Meanwhile, the redistributed rewards are denser and more timely, providing more informative feedback for policy optimization. It enables more accurate gradient estimation and improves both inference SLOs and regulation provision performance.

\subsection{Mean-Field Based Hierarchical Optimization}

Although the reformulated MDP with redistributed rewards provides dense feedback, optimizing the hierarchical scheduling framework remains challenging due to the curse of dimensionality. In particular, this challenge mainly arises from the lower-layer coordination among a large number of replicas, where the joint action space grows exponentially with the scale. To address this issue, we adopt a hierarchical optimization framework based on mean-field approximation. The upper layer focuses on coarse-grained request dispatching, while the lower layer handles fine-grained, replica-level resource allocation.

In the upper layer, the cluster scheduler is trained using deep Q learning (DQN) to dispatch incoming requests to the appropriate replicas. Its objective is to learn the optimal policy $\pi_u^*$ that maximizes the expected return: 
\begin{equation}
a_t^{u,*}
=
\arg\max_{a_t^u}
Q^{u,*}(s_t^u,a_t^u).
\end{equation}
where the optimal action value satisfies the Bellman equation:
\begin{equation}
\begin{aligned}
Q^{u,*}(s^u_t, a^u_t) = \mathbb{E}_{\substack{
    \hat{r}_t,\, s^u_{t+1} \sim \\
    \mathcal{P}(\cdot \mid s^u_t, a^u_t, \{a^l_{i,t}\})
}} \left[ \hat{r}_t + \gamma \max_{a^u_{t+1}} Q^{u,*}(s^u_{t+1}, a^u_{t+1}) \right].
\end{aligned}
\end{equation}
The network $Q^u(s^u_t,a^u_t;\theta^u)$ is optimized using the standard temporal-difference loss, while the environment transition $\mathcal{P}(\cdot\mid s^u_t, a^u_t, \{a^l_{i,t}\})$ implicitly depends on the collective lower-layer actions of all replicas.

In the lower layer, each LLM replica scheduler performs fine-grained token-level resource allocation. Let $\boldsymbol{s}_t^l=(s_{1,t}^l,\ldots,s_{N,t}^l)$ and $\boldsymbol{a}_t^l=(a_{1,t}^l,\ldots,a_{N,t}^l)$ denote the joint lower-layer state and action, respectively. Directly learning a centralized action-value function \(Q^l(\boldsymbol{s}_t^l,\boldsymbol{a}_t^l)\) is computationally intractable in large-scale systems because the joint action space is the Cartesian product of the individual action spaces and therefore grows combinatorially as the number of replicas increases. This also leads to an increasingly high-dimensional critic input and substantially increases the difficulty of coordinated exploration and value estimation. We therefore adopt a mean-field approximation that represents the interaction between replica $i$ and the remaining replicas through their mean action\(\bar{a}_{-i,t}^l=\frac{1}{N-1}\sum_{m\neq i}a_{m,t}^l.\) Accordingly, the centralized action-value function is approximated by a replica-specific mean-field action-value function \(Q_i^l(s_{i,t}^l,a_{i,t}^l,\bar{a}_{-i,t}^l)\), where $\bar{a}_{-i,t}^l$ summarizes the aggregate resource-allocation behavior of the other replicas and serves as a proxy for their aggregate power dynamics. The mean-field formulation replaces explicit dependence on the full joint action with a fixed-dimensional summary of the other replicas' actions. Computing this mean field scales linearly with the number of replicas, thereby alleviating the combinatorial growth of the joint action space and enabling scalable lower-layer coordination.

Each replica learns a local policy $\pi_i^l$ that maximizes its expected action value conditioned on the mean field:
\begin{equation}
\pi_i^{l,*}=\arg\max_{\pi_i^l}\mathbb{E}_{a_{i,t}^l\sim\pi_i^l}\left[Q_i^l\left(s_{i,t}^l,a_{i,t}^l,\bar{a}_{-i,t}^{l,*}\right)\right],
\end{equation}
where $\bar{a}_{-i,t}^{l,*}$ is induced by the stationary policies of the other replicas. For notational simplicity, define
\begin{equation}
\begin{aligned}
Q_{i,t}&\triangleq Q_i^l(s_{i,t}^l,a_{i,t}^l,\bar{a}_{-i,t}^l),\\
V_{i,t}&\triangleq\mathbb{E}_{a_{i,t}^l\sim\pi_i^l}\left[Q_i^l(s_{i,t}^l,a_{i,t}^l,\bar{a}_{-i,t}^l)\right].
\end{aligned}
\end{equation}

The local critic and actor networks are trained using PPO. The critic, parameterized by $\psi_i^l$, minimizes the temporal-difference loss
\begin{equation}
\mathcal{L}_i^{\mathrm{critic}}(\psi_i^l)=\left(Q_{i,t}-y_{i,t}\right)^2,
\qquad
y_{i,t}=\hat{r}_t+\gamma V_{i,t+1}.
\end{equation}

The actor, parameterized by $\theta_i^l$, is optimized by minimizing the negative clipped PPO surrogate objective: \begin{equation} \mathcal{L}_i^{\mathrm{actor}}(\theta_i^l) = -\mathbb{E}\left[ \min\left( \rho_{i,t}A_{i,t}, \operatorname{clip}(\rho_{i,t},1-\eta,1+\eta)A_{i,t} \right) \right], \end{equation} where \( \rho_{i,t} = \frac{\pi_i^l(a_{i,t}^l\mid s_{i,t}^l;\theta_i^l)} {\pi_{i,\mathrm{old}}^l(a_{i,t}^l\mid s_{i,t}^l)}, A_{i,t}=Q_{i,t}-V_{i,t}, \)and $\eta$ is the clipping parameter.

The mean-field formulation replaces the dependence on the full joint action with the mean action of the other replicas, reducing the interaction complexity from exponential to linear in the number of replicas and enabling scalable lower-layer coordination.

\textit{\textbf{Lemma 2 (Convergence of Mean-Field PPO):} }
\textit{
Consider an MDP with bounded rewards and a unique stationary mean-field Nash equilibrium $\pi^{l,\mathrm{NE}}.$ The mean action at decision step $t$ is defined as
$\bar{a}^l_{-i,t}
=
\frac{1}{N-1}\sum_{m\neq i}a^l_{m,t}$.
The critic is updated via (13), and the actor is updated via (14) with clipping parameter $\eta$. If the learning rate satisfy the Robbins–Monro conditions, then
\[
\lim_{q \to \infty} \theta_q^l = \theta^{l,\mathrm{NE}}, 
\quad 
\lim_{q \to \infty} \pi_{\theta_q^l} = \pi^{l,\mathrm{NE}},
\]
where $\theta^{l,\mathrm{NE}}$ denotes the PPO parameterization of the stationary mean-field Nash equilibrium policy profile $\pi^{l,\mathrm{NE}}$.
}
\begin{proof}
See Appendix~C for the complete proof.
\end{proof}

To further enhance the convergence properties of the hierarchical architecture, we adopt a two-time-scale update scheme. Specifically, the lower-layer replicas optimize token allocation at a fast pace, allowing the upper-layer scheduler to update more slowly based on an approximately stationary lower-level environment.

\textit{\textbf{Theorem 2 (Convergence of the Hierarchical Optimization Framework):}  
Let $\theta^u_q$ and $\theta^l_q$ denote the parameters of the upper-layer DQN and the lower-layer mean-field PPO, respectively. If the learning rates $\alpha_{l,q}$ and $\alpha_{u,q}$ satisfy the Robbins–Monro conditions and $\alpha_{u,q}/\alpha_{l,q} \to 0$, then the coupled iterates asymptotically converge to a stationary solution of the form
\begin{equation}
(\theta^u_q, \theta^l_q) \to \big( \theta^{u,*},\, \theta^{l,\mathrm{NE}}(\theta^{u,*}) \big),
\end{equation}
where $\theta^{l,\mathrm{NE}}(\theta^{u,*})$ corresponds to the stationary mean-field Nash equilibrium policy profile $\pi^{l,\mathrm{NE}}.$ induced by $\theta^{u,*}$.  }
\begin{proof}
See Appendix~D for the complete proof.
\end{proof}

\begin{figure}[htb]
\vspace{-3mm}
    \centering
    \includegraphics[width=8.8cm]{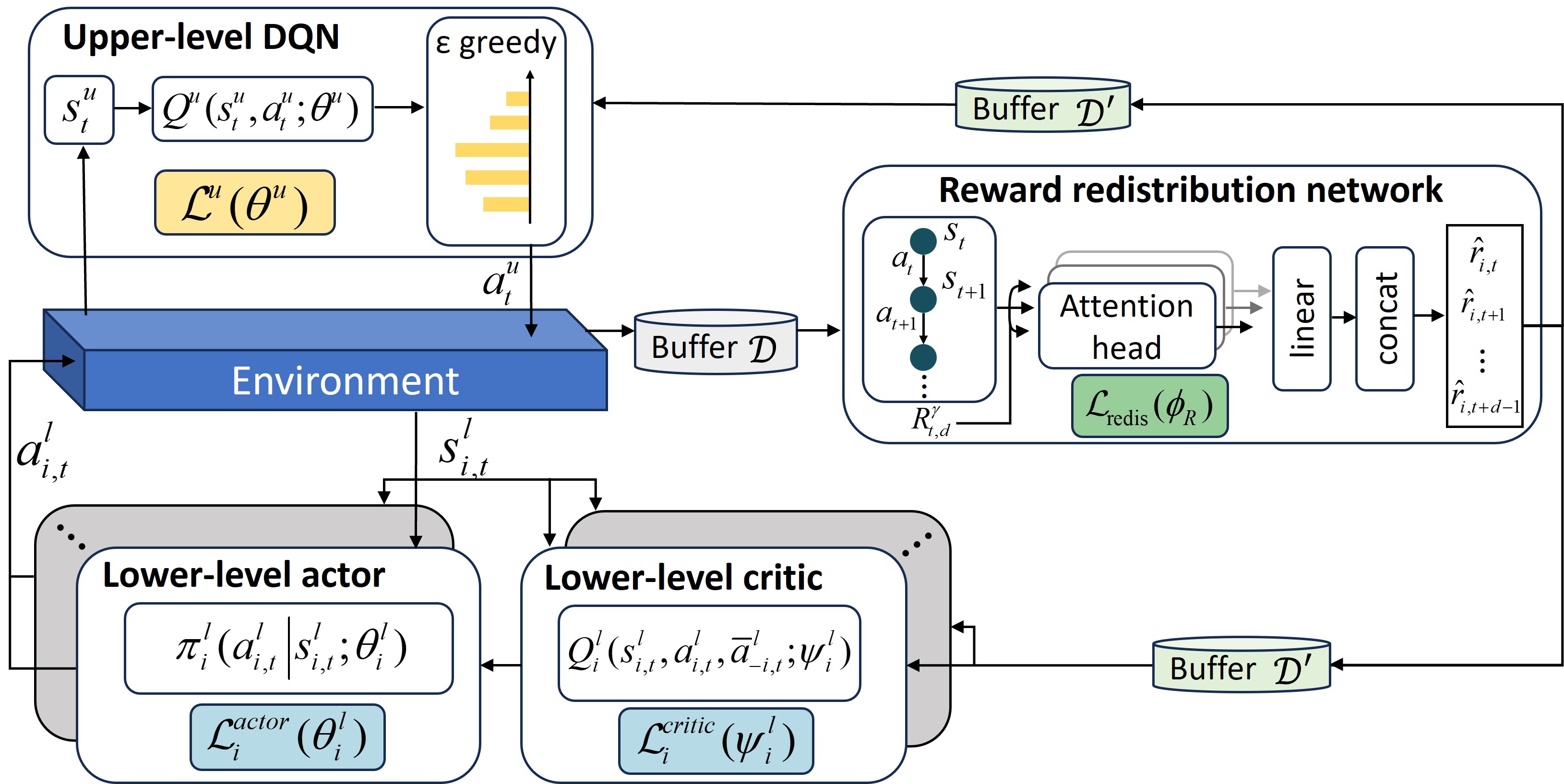}
    \caption{Overall workflow of the hierarchical optimization with reward redistribution.}
    \label{overall workflow}
\end{figure}

\begin{algorithm}[t]
\caption{Hierarchical Scheduling with Reward Redistribution}

\textbf{Initialize:} Episodes $N_{\mathrm{ep}}$, horizon $T$, replay buffer $\mathcal{D}$\;
\textbf{Initialize Networks:} $Q^u(\theta^u)$, $\{\pi_i^l(\theta_i^l),Q_i^l(\psi_i^l)\}_{i=1}^{N}$, and redistribution network $f_R(\phi_R)$ with random weights; 
\textbf{Parameters:} Update intervals $K_l,K_u$ ($K_u\gg K_l$), DQN exploration rate $\epsilon$, PPO clip $\eta$\;

\For{$\mathrm{episode}=1$ to $N_{\mathrm{ep}}$}{
    Reset the environment and obtain initial states\;
    \For{$t=1$ to $T$}{
        \textbf{Upper layer (DQN):} select $a_t^u$ using an $\epsilon$-greedy policy based on $Q^u$\;
        
        \textbf{Lower layer (mean-field PPO):} each replica $i$ samples $a_{i,t}^l$ from $\pi_i^l(\cdot\mid s_{i,t}^l;\theta_i^l)$\;
        
        Execute the joint actions, observe $r_t$ and the next states, and store the transition in $\mathcal{D}$\;
        
        \If{$t\,\%\,K_l=0$}{
            Update the redistribution network:
            \[
            \phi_R\leftarrow\phi_R-\alpha_R\nabla_{\phi_R}\mathcal{L}_{\mathrm{redis}}(\phi_R)
            \]
            Obtain $\{\hat r_\tau\}$ and replace the original rewards $\{r_\tau\}$ in the sampled segments\;
            
            \ForEach{replica $i$}{
                Update the mean-field critic:
                \[
                \psi_i^l\leftarrow\psi_i^l-\alpha_{l,c}\nabla_{\psi_i^l}\mathcal{L}_i^{\mathrm{critic}}(\psi_i^l)
                \]
                Update the local actor:
                \[
                \theta_i^l\leftarrow\theta_i^l-\alpha_{l,a}\nabla_{\theta_i^l}\mathcal{L}_i^{\mathrm{actor}}(\theta_i^l)
                \]
                Set $\theta_{i,\mathrm{old}}^l\leftarrow\theta_i^l$\;
            }
        }
        
        \If{$t\,\%\,K_u=0$}{
            Update the upper-layer DQN:
            \[
            \theta^u\leftarrow\theta^u-\alpha_u\nabla_{\theta^u}\mathcal{L}^u(\theta^u)
            \]
        }
    }
}
\end{algorithm}

To summarize, the proposed method adopts a two-stage scheduling framework for LLM inference requests, as illustrated in Fig.~\ref{overall workflow}. The complete training and update procedure is presented in Algorithm~1. Once training converges, online scheduling requires only forward inference of the upper-layer and lower-layer network, without reward redistribution.

\section{Case Study}

\subsection{Experimental Setting}
We consider a cluster equipped with 10,000 H100 GPUs, with 312 inference replicas deployed. The inference execution process is simulated via VIDUR~\cite{agrawal2024vidur}, which models the response latency and model FLOPs utilization (MFU) under different scheduling policies. Based on the execution state simulated at each scheduling step, we estimate its power consumption as
\(
p_{i,t}
=
\mathrm{MFU}_{i,t}\times P_i^{\mathrm{rated}},
\)
where $P_i^{\mathrm{rated}}$ denotes the rated power of the GPUs assigned to that replica. Inference requests are generated from a real-world data traces \cite{MicrosoftAzure_AzureLLMInferenceTraceConv} with an average arrival rate of 10,000 requests per second. The baseline power of the cluster is set to 60\% of its rated capacity, and the regulation capacity $R_{\mathrm{fre}}$ is set to 10\%. The PJM regulation service is adopted as a representative case study, and the regulation signal is obtained from the PJM RegD dataset~\cite{pjm2025ancillary}.

In addition to the average tracking error, regulation performance is evaluated using a PJM-based score. Since each regulation interval $k$ lasts 2~s, the requested and delivered responses over the 
$g$-th 10-s scoring interval are calculated as
\begin{equation}
\begin{aligned}
P_g^{\mathrm{req}}
&=
\frac{1}{5}
\sum_{k=5g-4}^{5g}
\left(P_k^{\mathrm{tar}}-P^{\mathrm{base}}\right),\\
P_g^{\mathrm{del}}
&=
\frac{1}{5}
\sum_{k=5g-4}^{5g}
\left(\bar{P}_k-P^{\mathrm{base}}\right),
\end{aligned}
\label{eq:ten_second_response}
\end{equation}
where $P^{\mathrm{base}}$ is the baseline cluster power. For each 5-min 
settlement interval $c$, the PJM-based performance score ~\cite{pjmManual12} is calculated as 
\begin{equation}
\mathrm{Score}_{c}^{\mathrm{PJM}}
=
1-
\frac{
\displaystyle
\frac{1}{30}
\sum_{g\in\mathcal{G}_c}
\left|P_g^{\mathrm{del}}-P_g^{\mathrm{req}}\right|
}{
\displaystyle
\frac{1}{60}
\sum_{g\in\mathcal{G}_c}
\left|P_g^{\mathrm{req}}\right|
+
0.5R_{\mathrm{fre}}P^{\mathrm{rated}}
},
\label{eq:pjm_performance_score}
\end{equation}
where $\mathcal{G}_c$ contains the 30 scoring intervals within 
settlement interval $c$.

The key parameters of the proposed framework are summarized in Table~\ref{tab:implementation_params}.
\begin{table}[h]
\caption{Key implementation parameters of the proposed framework.}
\label{tab:implementation_params}
\centering
\renewcommand{\arraystretch}{1.1}
\begin{tabular}{lc}
\hline
\textbf{Parameter} & \textbf{Value} \\
\hline
Regulation interval & 2 s \\
Scheduling decision interval & 50 ms \\
Reward segment length \(d\) & 40 \\
Reward weights \((w_{\mathrm{fre}},w_{\mathrm{SLO}},w_{\mathrm{com}})\)
& \((1,10^{-4},5\times10^{-8})\) \\
\hline
\end{tabular}
\end{table}

\vspace{-2mm}
\subsection{LLM Inference Scheduling for Frequency Regulation}

We first evaluate the effectiveness of the proposed method in providing regulation services and the associated scheduling details, as shown in Figs.~5 and~6.

\begin{figure}[htb]
 \vspace{-3mm}
    \centering
    \includegraphics[width=8.5cm]{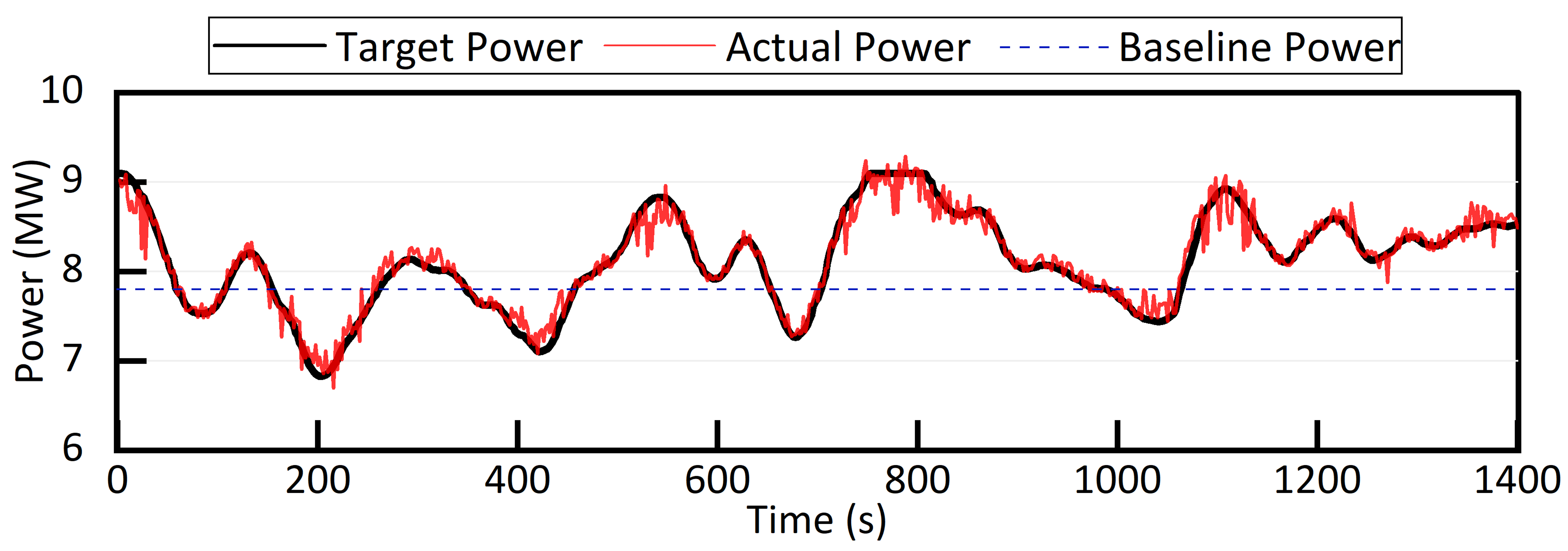}
        \vspace{-3mm}
    \caption{Frequency regulation performance of an LLM inference cluster under a 10\% regulation capacity.}
    \label{basecase}
\end{figure}

Figure~\ref{basecase} presents the power response at the 2-s regulation signal resolution under a 10\% regulation capacity. Although the cluster power generally follows the target signal, pointwise fluctuations occur because inference requests arrive stochastically and are executed as discrete workloads with heterogeneous processing times and power demands. Moreover, the TTFT and E2E constraints limit the extent to which workload execution can be shifted solely for power tracking. Quantitatively, the proposed method achieves an average tracking error of 1.17\%, an average TTFT of 0.94~s, and an average E2E latency of 6.36~s. Under the PJM scoring rule, it achieves a mean performance score of 0.926 and a minimum 5-min score of 0.910. Therefore, despite the visible fluctuations, the response maintains consistent regulation delivery over the evaluated intervals.

\begin{figure}[htb]
    \centering
    \includegraphics[width=8.8cm]{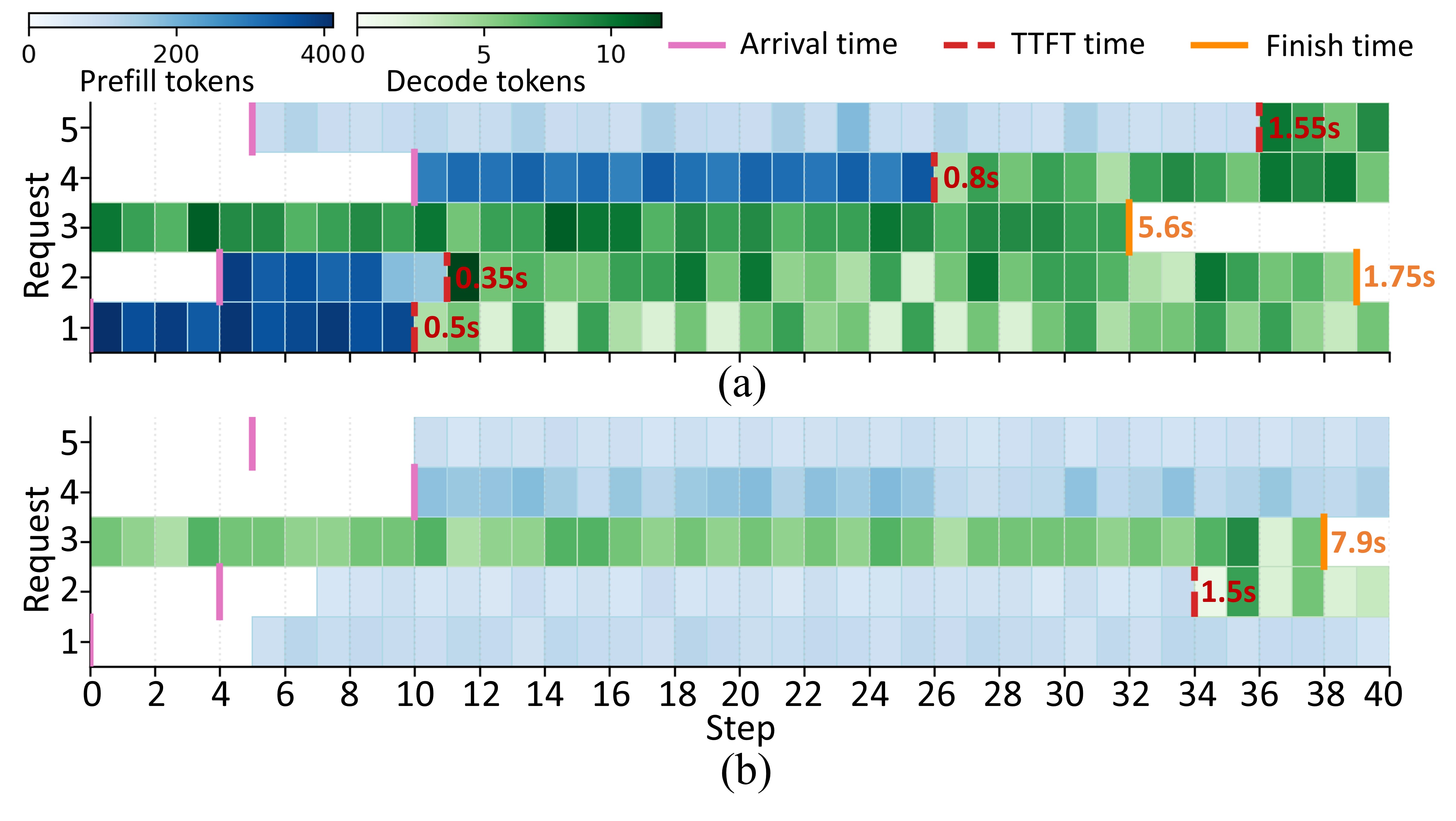}
    \vspace{-8mm}
    \caption{Scheduling details of (a) the throughput-maximization strategy and (b) the proposed strategy.}
    \label{baselinecomp}
\end{figure}

Figure~\ref{baselinecomp} shows scheduling decisions for five requests from one replica over 40 steps during the up-regulation interval, corresponding to $t\in [400, 402]$ in Fig.~5. The baseline strategy, as illustrated in Fig. 6(a), executes requests immediately to maximize throughput. In contrast, the proposed method, as illustrated in Fig. 6(b), strategically defers the request execution and limits the number of processed tokens to reduce power and follow the target regulation signal. Although it leads to a slight increase in TTFT and E2E latency, the proposed method effectively tracks regulation signals while keeping the SLOs within acceptable limits.

\subsection{Impact of Reward Redistribution}

To validate the effectiveness of reward redistribution, we compare the proposed method with a baseline where rewards are only issued at the end of each regulation interval and no reward redistribution is applied.

\begin{figure}[htb]
    \centering
    \includegraphics[width=8.5cm]{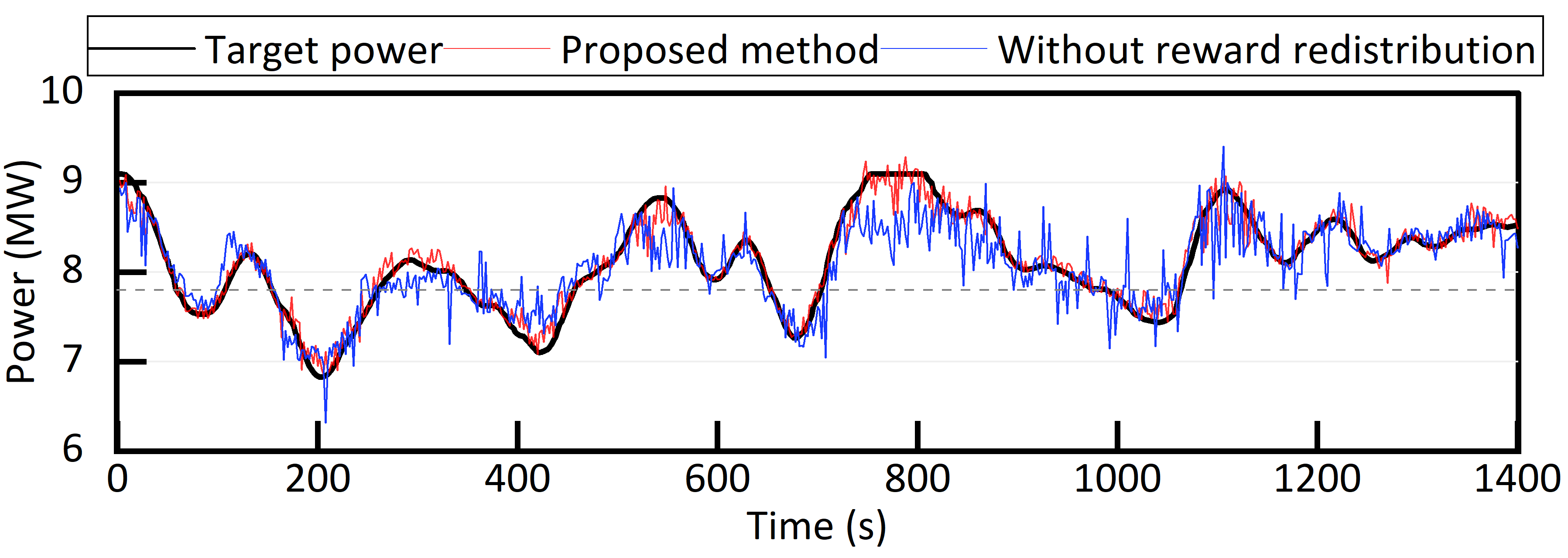}
    \caption{Comparison of regulation service with and without reward redistribution.}
    \label{without}
\end{figure}

\begin{table}[htb]
    \centering
    \caption{Regulation Performance With and Without Reward Redistribution}
    \label{tab:redistribution_score}
    \renewcommand{\arraystretch}{1.05}
    \begin{tabular*}{\columnwidth}{@{\extracolsep{\fill}}lcc}
        \toprule
        Method & Mean PJM Score & Min. PJM Score \\
        \midrule
        Without reward redistribution & 0.799 & 0.739 \\
        Proposed method & \textbf{0.926} & \textbf{0.910} \\
        \bottomrule
    \end{tabular*}
\end{table}

As shown in Fig.~\ref{without}, the method without reward redistribution exhibits unstable tracking performance, characterized by sharp spikes in the power curve and frequent deviations from the target power. This instability arises because delayed and coupled feedback prevents the learning process from extracting effective policy-improvement signals. In contrast, the proposed method enables the cluster power to closely follow the regulation signals. This improvement is attributed to redistributing delayed rewards to individual scheduling decisions, which provides denser learning signals and enables more effective policy updates for regulation. Table~\ref{tab:redistribution_score} further compares the two methods using the PJM-based performance metric. Without reward redistribution, the minimum performance score of 0.739 falls below the 0.75 threshold used in PJM regulation-resource qualification tests. In contrast, the proposed method achieves both mean and minimum scores above 0.90, demonstrating more accurate and consistent tracking performance.

\begin{figure}[htb]
    \centering
    \includegraphics[width=8.5cm]{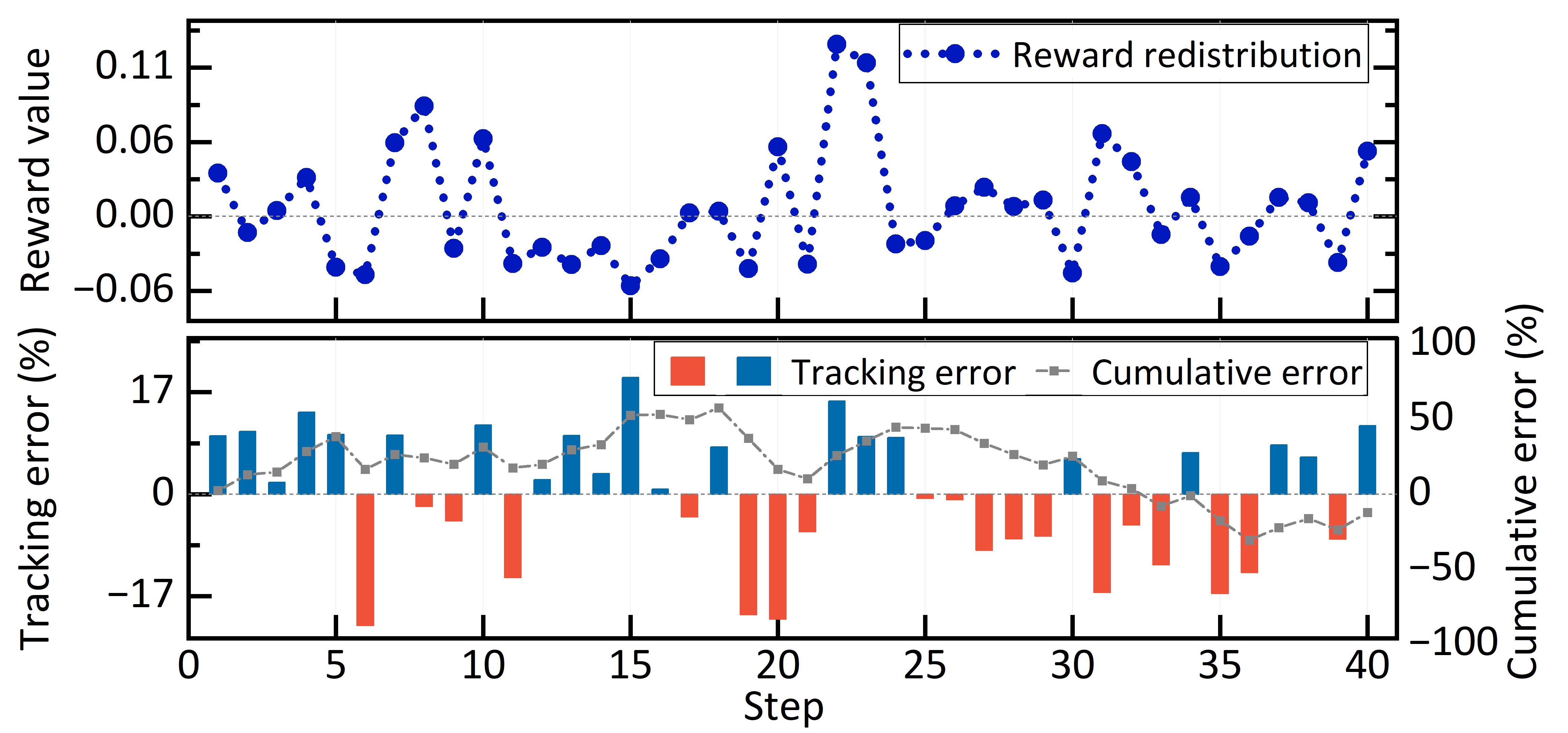}%
    \caption{Reward redistribution details within one regulation interval using the proposed method.}
    \label{detailed}
\end{figure}

Figure ~\ref{detailed} illustrates the reward redistribution process within a single regulation interval. Although instantaneous tracking errors can reach 90\%, the cumulative error remains negligible by the end of the interval. It indicates that our method prioritizes interval-level consistency over immediate deviations. For instance, the action at step 20 receives a positive reward for correcting accumulated errors despite a large local deviation. This mechanism allows for flexible execution adjustments while ensuring overall regulation performance.

To further evaluate the impact of reward redistribution, we design three comparative methods with different reward assignment strategies within the same scheduling framework.

\textbf{Case I (Without Reward Redistribution):}
No reward redistribution is applied. 

\textbf{Case II (Uniform Redistribution):}
Interval-level rewards are evenly distributed across all steps within the interval.

\textbf{Case III (RUDDER Redistribution):}
Rewards are redistributed along the trajectory using an LSTM-based return decomposition network \cite{arjona2019rudder}.

\begin{figure}[htb]
    \centering
    \includegraphics[width=7.8cm]{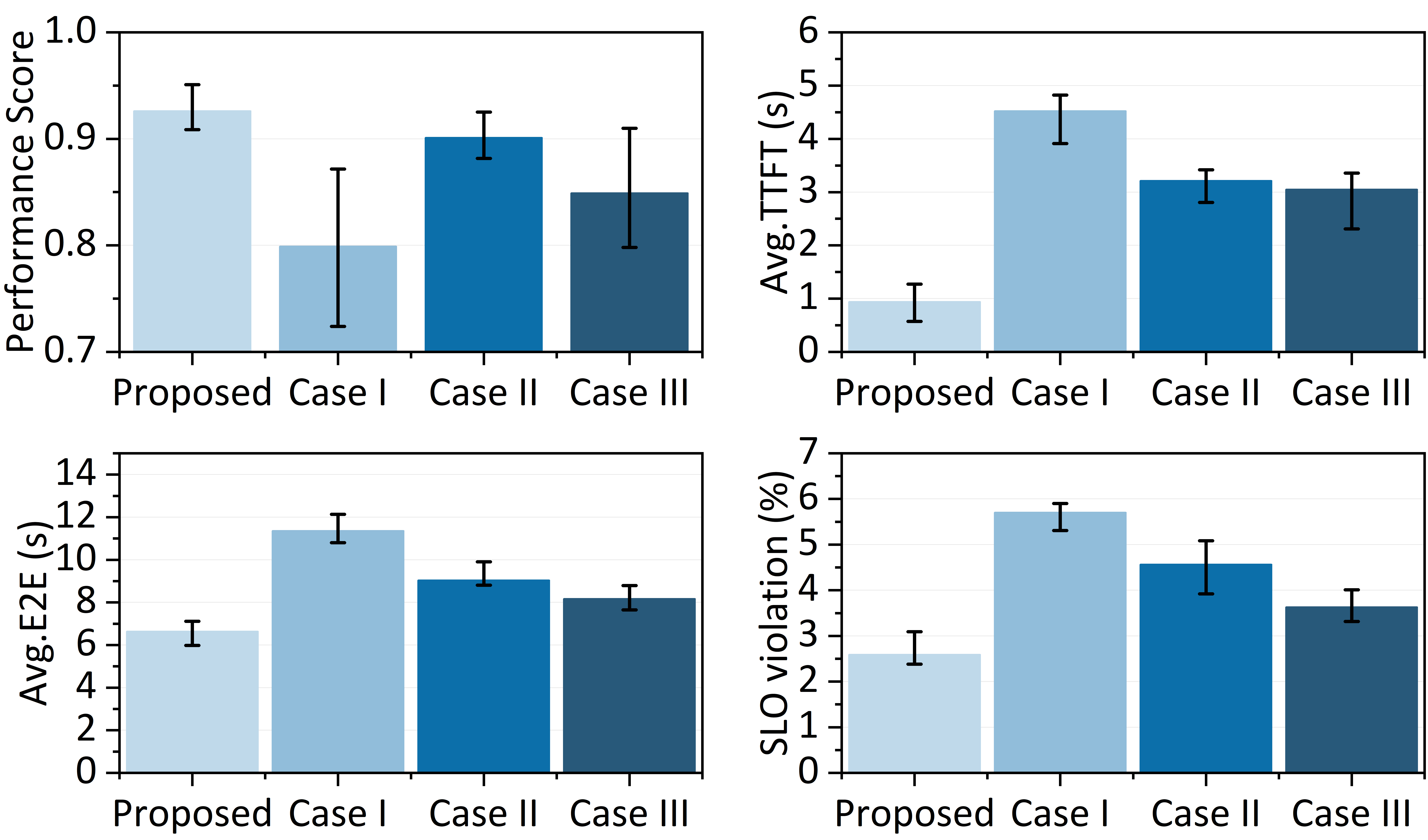}
    \caption{Performance comparison of different reward redistribution methods.}
    \label{credit_compare}
    \vspace{-5mm}
\end{figure}

\begin{table*}[t]
\centering
\caption{Performance Comparison under Different Replica Scales}
\begin{tabular}{ccccccc}
\hline
Number of replicas & Method & Training time (h) & Avg. MAE & Avg. TTFT (s) & Avg. E2E (s) & SLO vio.\\
\hline
\multirow{3}{*}{5}   
 & Fully Centralized   & 2.1  & \textbf{0.72\%} & \textbf{0.61} & \textbf{3.93} & \textbf{1.06\%}  \\
 & Mean-Field based   & 2.0  & 0.80\% & 0.67 & 4.16 & 1.32\%  \\
 & Fully Decentralized & 1.7 & 1.05\% & 0.81 & 4.38 & 1.71\%  \\
\hline
\multirow{3}{*}{10}  
 & Fully Centralized   & 6.4  & \textbf{0.98\%} & 0.75 & \textbf{5.01 }& \textbf{1.15\%}  \\
 & Mean-Field based   & 4.9  & 1.01\% & \textbf{0.73} & 5.22 & 1.58\%  \\
 & Fully Decentralized & 3.7  & 1.37\% & 1.31 & 5.51 & 2.16\%  \\
\hline
\multirow{3}{*}{20}  
 & Fully Centralized   & 18.5  & 1.03\% & \textbf{0.82} & \textbf{5.17} & \textbf{1.78\%}  \\
 & Mean-Field based  & 9.1  & \textbf{1.02\%} & 0.88 & 5.39 & 1.92\%  \\
 &Fully Decentralized & 7.2  & 2.44\% & 1.49 & 7.02 & 3.08\%  \\
\hline
\multirow{3}{*}{50}  
 & Fully Centralized   & --   & --   & --   & --   & --   \\
 & Mean-Field  based   & 16.8  & \textbf{1.12\%} & \textbf{1.04} & \textbf{5.98} & \textbf{1.74\%}  \\
 &Fully Decentralized & 15.4  & 4.13\% & 2.58 & 9.40 & 5.46\%  \\
\hline
\multirow{3}{*}{100} 
 &Fully Centralized   & --   & --   & --   & --   & --   \\
 & Mean-Field  based   & 21.6 & \textbf{1.34\%} & \textbf{1.22} & \textbf{6.06} & \textbf{1.97\%}  \\
 &Fully Decentralized & 19.3  & 6.22\% & 5.76 & 13.82 & 8.03\% \\
\hline
\multirow{3}{*}{200} 
 &Fully Centralized   & --   & --   & --   & --   & --   \\
 & Mean-Field based    & 36.7 & \textbf{1.19\%} & \textbf{1.03} & \textbf{6.13}& \textbf{2.30\%} \\
 &Fully Decentralized & 35.5 & 7.75\% & 7.05 & 18.51 & 12.40\% \\
\hline
\end{tabular}
\vspace{-3mm}
\label{scalability}
\end{table*}

Figure~\ref{credit_compare} compares the four methods in terms of the PJM-based performance score and inference service quality. Case~I suffers from sparse interval-level feedback, resulting in the lowest average score and substantial performance variation. Case~II improves regulation tracking through uniform reward redistribution, but assigning identical rewards to all actions fails to distinguish their individual contributions, leading to higher TTFT, E2E latency, and SLO violations. Case~III improves credit assignment through RUDDER-based return decomposition, but is less effective in capturing the coupled effects of scheduling decisions under time-varying requests and regulation signals. In contrast, the proposed method accounts for both long-term action impacts and temporal regulation dynamics, enabling more accurate reward redistribution. Consequently, it achieves improved tracking performance while maintaining stable SLOs.

\begin{figure}[h]
    \centering
    \includegraphics[width=8cm]{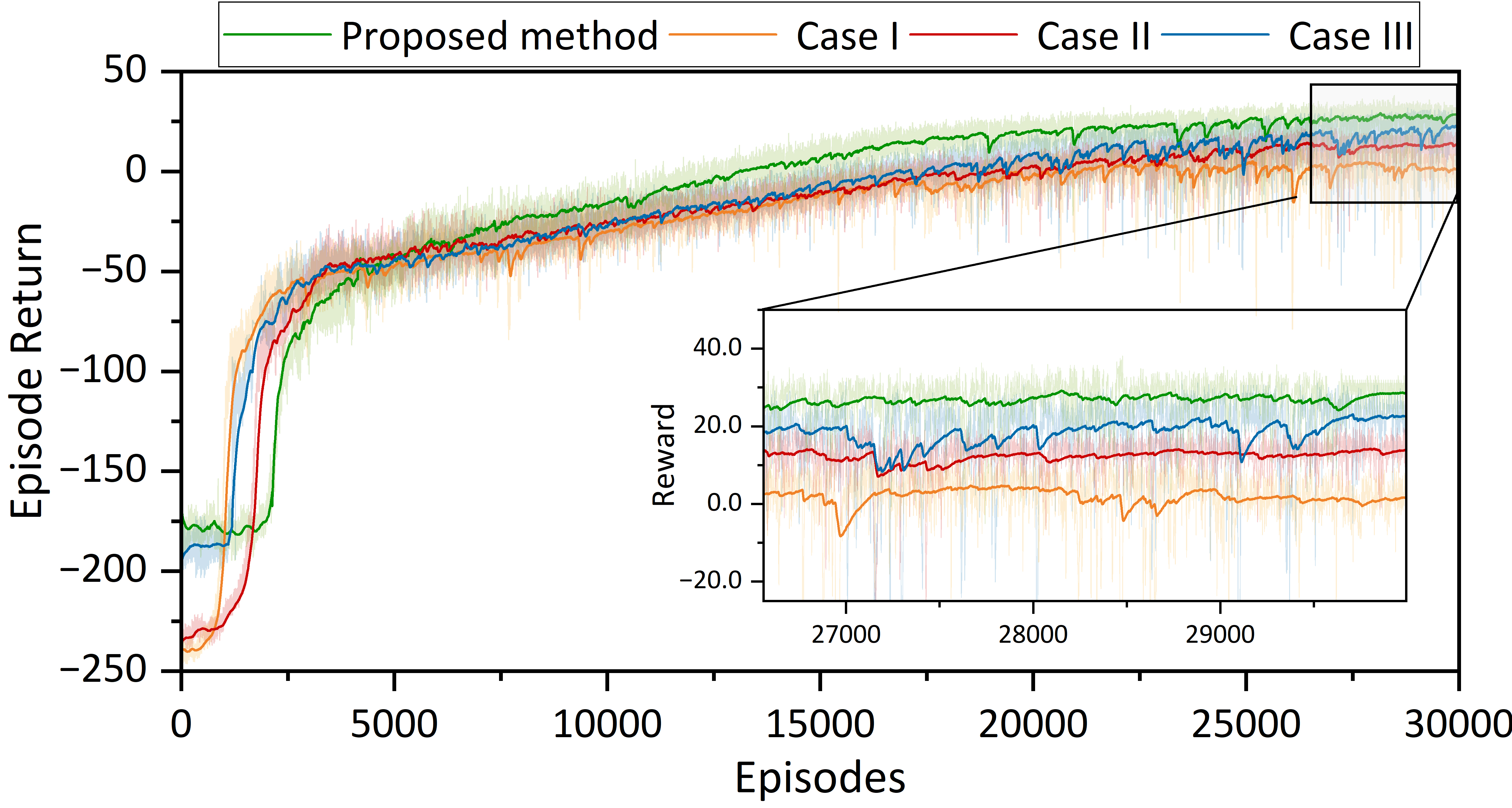}
    \caption{Comparison of training performance across different reward redistribution methods.}
    \label{creditassignmenttrain}
\end{figure}

As shown in Fig.~\ref{creditassignmenttrain}, the proposed method achieves competitive performance with reduced variance compared to the baselines. Whereas standard approaches face challenges with convergence under sparse feedback, our method shows steady improvement, which validates the effectiveness of the reward redistribution strategy in stabilizing the training process.

\vspace{-3mm}
\subsection{Scalability Performance}

In this subsection, we compare training efficiency and performance under increasing numbers of replicas, where the lower-layer scheduling adopts fully centralized, fully decentralized, and the proposed mean-field based methods.

\textbf{Fully Centralized:}  
Each replica maintains an individual actor, but all actors are trained with the global state and joint actions, resulting in fully centralized training and execution.

\textbf{Fully Decentralized:}  
Each replica trains its own policy using local interactions. States of other replicas are observable, but their actions are not accessible.

Table~\ref{scalability} summarizes the performance as the number of replicas increases. With a small number of replicas, the centralized method achieves the best tracking accuracy and lowest latency by exploiting global information. The proposed mean-field method attains comparable performance with substantially shorter training time, and in some cases outperforms the centralized baseline. The decentralized method converges fastest but relies solely on local observations, resulting in limited coordination and degraded performance.

As the system scales, the centralized method incurs rapidly growing computational overhead and fails to converge at 50 replicas under the same hardware configuration. Meanwhile, the decentralized method suffers from severe non-stationarity and lack of coordinated behavior, leading to degraded tracking accuracy and higher latency. In contrast, the proposed mean-field method maintains stable performance and moderate training time across scales. Although its accuracy slightly decreases under large settings, the degradation remains acceptable, which demonstrates scalability and robustness. Overall, the proposed approach offers an effective balance between coordination efficiency and computational cost for large-scale LLM inference scheduling.

\vspace{-3mm}
\subsection{Regulation Capacities and Performance Deterioration}

\begin{figure}[h]
    \centering
    \includegraphics[width=\linewidth]{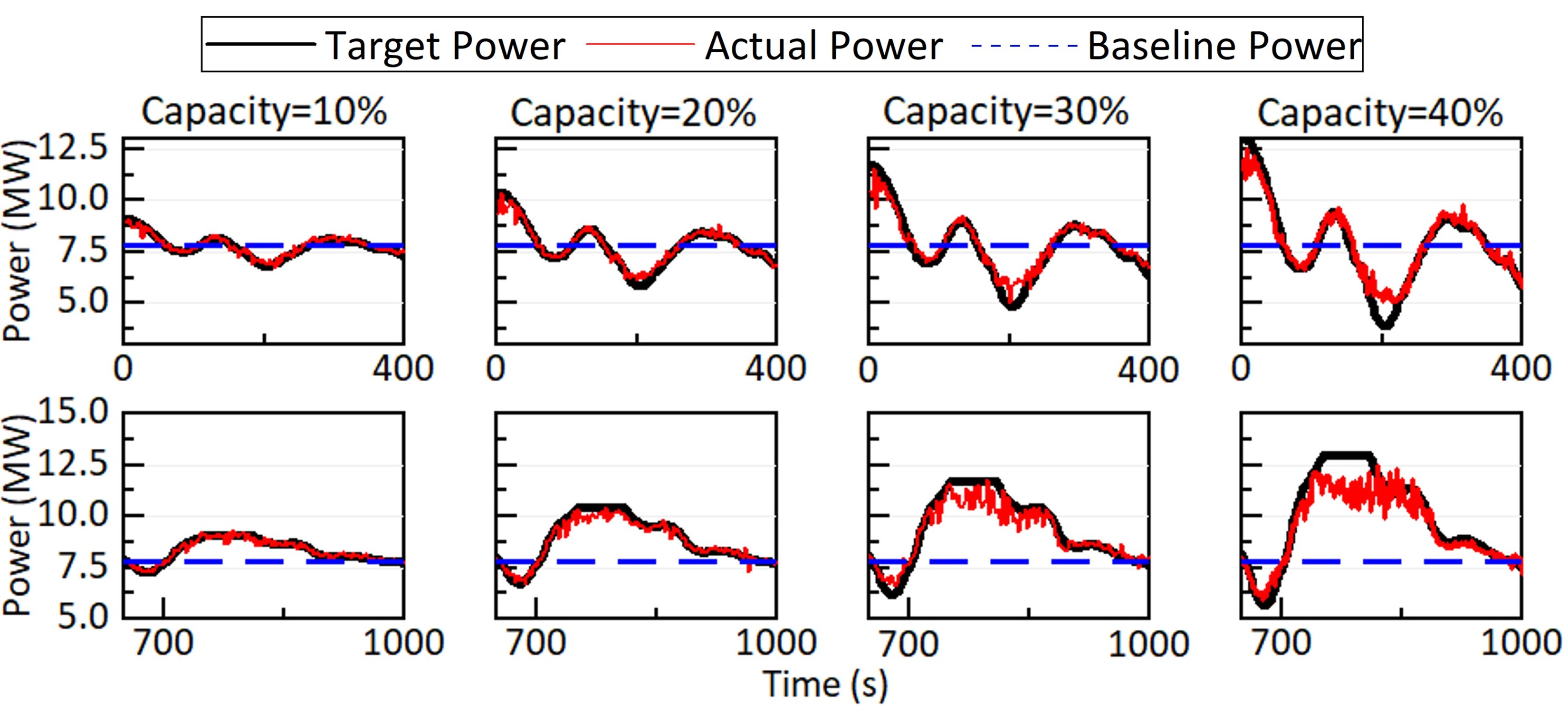}%
    \vspace{-3mm}
    \caption{Tracking performance under different regulation capacities.}
    \label{jubu}
\end{figure}

We evaluate the proposed method under regulation capacities $R_{\mathrm{fre}}$ from 10\% to 40\%, with results summarized in Table~\ref{reg_capacity} and Fig.~\ref{jubu}.

Figure~\ref{jubu} illustrates the local tracking performance under different regulation capacities. As the regulation capacity increases, more inference requests must be deferred during upward regulation to reduce power consumption. However, SLOs limit the allowable delay, making perfect tracking infeasible at high capacities. In addition, prolonged downward regulation is constrained by request arrivals, further increasing tracking deviations. These results indicate that practical LLM inference scheduling requires a careful trade-off among regulation capacity, tracking performance, and SLOs. 

\begin{table}[htb]
    \centering
    \vspace{-3mm}
    \caption{Performance under Different Regulation Capacities}
    \label{reg_capacity}
    \renewcommand{\arraystretch}{1.05}
    \begin{tabular*}{\columnwidth}{@{\extracolsep{\fill}}cccccc}
        \toprule
        Capacity 
        & Avg. MAE
        & \shortstack{PJM Score\\(Mean/Min.)}
        & \shortstack{TTFT\\(s)}
        & \shortstack{E2E\\(s)}
        & \shortstack{SLO\\Vio.} \\
        \midrule
        10\% & 1.17\% & 0.926 / 0.910 & 0.94 & 6.36  & 2.59\% \\
        20\% & 1.69\% & 0.933 / 0.926 & 1.87 & 7.69  & 3.03\% \\
        30\% & 3.51\% & 0.880 / 0.859 & 2.48 & 10.13 & 3.86\% \\
        40\% & 5.07\% & 0.884 / 0.837 & 5.66 & 18.02 & 7.41\% \\
        \bottomrule
    \end{tabular*}
\end{table}

As shown in Table~\ref{reg_capacity}, all evaluated 5-min performance scores remain above 0.75. However, increasing the regulation capacity requires greater adjustment of inference execution, resulting in longer TTFT and E2E latency. At capacities of 30\% and above, the increased tracking errors and SLO violations indicate that practical capacity commitments should balance regulation performance against inference service quality.

\section{Conclusion}
This paper explores the potential of the LLM inference cluster to provide frequency regulation service. We formulate the problem as a hierarchical MDP that integrates request dispatching and token-level resource allocation. To mitigate the asynchrony between scheduling decisions and regulation signals, a reward redistribution mechanism is introduced to ensure dense feedback. A mean-field approximation based optimization framework is further developed to coordinate large numbers of replicas. Experiments with real-world data demonstrate that the proposed method achieves high tracking accuracy while maintaining SLOs, enabling the cluster to provide frequency regulation service through the flexibility offered by LLM inference request scheduling. 

Future work can extend the proposed framework in three directions. First, it can be adapted to sub-second frequency-response products, such as the UK Dynamic Containment service. Second, regulation-capacity bidding can be optimized to determine the available capacity under time-varying workloads and service requirements, while adaptive baseline estimation under changing operating conditions also warrants further investigation. Finally, workload-level scheduling can be coordinated with hardware-level controls, such as GPU power capping and DVFS, to exploit their complementary flexibility for more accurate and reliable regulationprovision.

\appendices

\section{}

\begin{proof}
Let \(\mathcal{B}=\{B_{t_b,d_b}\}_{b=1}^{N_{\mathrm{seg}}}\) be a non-overlapping partition of the trajectory, where \(d_b\in\mathbb{Z}_{>0}\) denotes the number of decision steps in segment \(b\). Each decision step belongs to exactly one segment, and the segments jointly cover the entire trajectory, i.e.,
\(\bigcup_{b=1}^{N_{\mathrm{seg}}}\{t_b,\ldots,t_b+d_b-1\}=\{1,2,\ldots,T\}\), and
\(\{t_b,\ldots,t_b+d_b-1\}\cap
\{t_{b'},\ldots,t_{b'}+d_{b'}-1\}=\emptyset\)
for \(b\neq b'\).

For each segment, define its discounted return as
\(R_{t_b,d_b}^{\gamma}
=\sum_{\tau=t_b}^{t_b+d_b-1}\gamma^{\tau-t_b}r_\tau\).
If the redistributed rewards satisfy
\(\sum_{\tau=t_b}^{t_b+d_b-1}\gamma^{\tau-t_b}\hat r_\tau
=R_{t_b,d_b}^{\gamma}\)
for every segment, then, for any trajectory,
\begin{align}
\sum_{\tau=1}^{T}\gamma^{\tau-1}\hat r_\tau
&=
\sum_{b=1}^{N_{\mathrm{seg}}}\gamma^{t_b-1}
\sum_{\tau=t_b}^{t_b+d_b-1}
\gamma^{\tau-t_b}\hat r_\tau \nonumber\\
&=
\sum_{b=1}^{N_{\mathrm{seg}}}\gamma^{t_b-1}R_{t_b,d_b}^{\gamma}
=
\sum_{\tau=1}^{T}\gamma^{\tau-1}r_\tau.
\end{align}

Taking the expectation under any policy \(\pi\) and initial state \(s\) yields
\begin{equation}
V^\pi_{\hat{\mathcal{M}}}(s)
=
\mathbb{E}_{\pi}
\left[
\sum_{\tau=1}^{T}\gamma^{\tau-1}\hat r_\tau
\,\middle|\,s
\right]
=
\mathbb{E}_{\pi}
\left[
\sum_{\tau=1}^{T}\gamma^{\tau-1}r_\tau
\,\middle|\,s
\right]
=
V^\pi_{\mathcal{M}}(s).
\end{equation}
Hence, every policy has the same value in \(\mathcal{M}\) and \(\hat{\mathcal{M}}\), and their optimal policy sets are identical:
\begin{equation}
\Pi^{*}_{\hat{\mathcal{M}}}
=
\bigcap_{s\in\mathcal{S}}
\arg\max_{\pi}V^\pi_{\hat{\mathcal{M}}}(s)
=
\bigcap_{s\in\mathcal{S}}
\arg\max_{\pi}V^\pi_{\mathcal{M}}(s)
=
\Pi^{*}_{\mathcal{M}}.
\end{equation}
Therefore, \(\mathcal{M}\) and \(\hat{\mathcal{M}}\) are decision-equivalent and admit the same set of optimal policies.
\end{proof}

\section{}

\begin{proof}
For each segment \(B_{t_b,d_b}\), define the discounted return-conservation error as
\(
\varepsilon_{t_b,d_b}
\triangleq
\sum_{\tau=t_b}^{t_b+d_b-1}
\gamma^{\tau-t_b}\hat r_\tau
-
R_{t_b,d_b}^{\gamma}.
\)
Since the segments are non-overlapping and jointly cover the trajectory, the difference between the original and redistributed discounted returns is
\begin{equation}
\left|
\sum_{\tau=1}^{T}\gamma^{\tau-1}\hat r_\tau
-
\sum_{\tau=1}^{T}\gamma^{\tau-1}r_\tau
\right|
=
\left|
\sum_{b=1}^{N_{\mathrm{seg}}}
\gamma^{t_b-1}\varepsilon_{t_b,d_b}
\right|.
\end{equation}
Therefore, the corresponding value-function gap satisfies
\begin{align}
\left|
V^\pi_{\mathrm{orig}}(s)
-
V^\pi_{\mathrm{redis}}(s)
\right|
&=
\left|
\mathbb{E}_{\pi}
\left[
\sum_{b=1}^{N_{\mathrm{seg}}}
\gamma^{t_b-1}\varepsilon_{t_b,d_b}
\right]
\right| \nonumber\\
&\leq
\mathbb{E}_{\pi}
\left[
\sum_{b=1}^{N_{\mathrm{seg}}}
\gamma^{t_b-1}
\left|\varepsilon_{t_b,d_b}\right|
\right].
\end{align}

For the \(N_{\mathrm{seg}}\) segments constituting the trajectory, define the average segment-wise MSE after \(n\) parameter updates as
\begin{equation}
\mathcal{L}_{\mathrm{redis}}^{(n)}
=
\frac{1}{N_{\mathrm{seg}}}
\sum_{b=1}^{N_{\mathrm{seg}}}
\left|\varepsilon_{t_b,d_b}\right|^2.
\end{equation}
To characterize how the decay of the redistribution loss affects the value-function gap, we make the following explicit sublinear loss-decay assumption, motivated by the convergence behavior commonly considered in analyses of stochastic first-order optimizers \cite{appendixb1,appendixb}:
\begin{equation}
\mathcal{L}_{\mathrm{redis}}^{(n)}
\leq
\frac{\mathcal{L}_{\mathrm{redis}}^{(0)}}{n^{1/2}},
\qquad
C\triangleq
\sqrt{\mathcal{L}_{\mathrm{redis}}^{(0)}}.
\end{equation}
Here, \(n\) denotes the number of parameter updates, and \(C\) is a finite constant determined by the initial empirical loss. By the Cauchy--Schwarz inequality, the discounted sum of the segment-wise redistribution errors satisfies
\begin{align}
\sum_{b=1}^{N_{\mathrm{seg}}}
\gamma^{t_b-1}
\left|\varepsilon_{t_b,d_b}\right|
&\leq
\left(
\sum_{b=1}^{N_{\mathrm{seg}}}\gamma^{2(t_b-1)}
\right)^{1/2}
\left(
\sum_{b=1}^{N_{\mathrm{seg}}}
\left|\varepsilon_{t_b,d_b}\right|^2
\right)^{1/2} \nonumber\\
&=
\left(
\sum_{b=1}^{N_{\mathrm{seg}}}\gamma^{2(t_b-1)}
\right)^{1/2}
\sqrt{N_{\mathrm{seg}}\mathcal{L}_{\mathrm{redis}}^{(n)}} \nonumber\\
&\leq
C\sqrt{N_{\mathrm{seg}}}\,n^{-1/4}
\left(
\sum_{b=1}^{N_{\mathrm{seg}}}\gamma^{2(t_b-1)}
\right)^{1/2}.
\end{align}
Since \(0<\gamma\leq1\), it follows that
\begin{equation}
\sum_{b=1}^{N_{\mathrm{seg}}}\gamma^{2(t_b-1)}
\leq
N_{\mathrm{seg}}.
\end{equation}
Substituting the resulting trajectory-wise bound into the value-function inequality gives
\begin{align}
\left|
V^\pi_{\mathrm{orig}}(s)
-
V^\pi_{\mathrm{redis}}(s)
\right|
&\leq
C\sqrt{N_{\mathrm{seg}}}\,n^{-1/4}
\left(
\sum_{b=1}^{N_{\mathrm{seg}}}\gamma^{2(t_b-1)}
\right)^{1/2} \nonumber\\
&\leq
C N_{\mathrm{seg}}n^{-1/4}.
\end{align}
\end{proof}

\section{}
\begin{proof}
In mean-field PPO, the critic corresponds to a mean-field action-value function $Q_i^l(s_{i,t}^l,a_{i,t}^l,\bar{a}_{-i,t}^l)$, whose convergence to the stationary mean-field Nash equilibrium $\pi^{l,\mathrm{NE}}$ has been shown in~\cite{yang2018mean}. Hence, the critic yields an asymptotically accurate estimate of the advantage function as the learning process approaches a stationary mean-field equilibrium.

Given the converged critic, the actor is updated by the clipped PPO rule:
\begin{equation}
\theta_{q+1}^l
=
\theta_q^l+\alpha_q\widehat{g}_q(\theta_q^l),
\end{equation}
where $\theta_q^l$ collectively denotes the parameters of all lower-layer actors at iteration $q$, $\{\alpha_q\}$ satisfies the Robbins--Monro conditions, and
\begin{equation}
\widehat{g}_q(\theta_q^l)
\triangleq
-\widehat{\nabla}_{\theta_q^l}
\mathcal{L}^{\mathrm{actor}}(\theta_q^l)
\end{equation}
denotes the negative stochastic gradient of the actor loss estimated from sampled trajectories at iteration $q$.

Accordingly, the mean update direction is
\begin{equation}
g(\theta_q^l)
\triangleq
\mathbb{E}\!\left[\widehat{g}_q(\theta_q^l)\right]
=
-\nabla_{\theta_q^l}\mathcal{L}^{\mathrm{actor}}(\theta_q^l).
\end{equation}

Since the gradient estimates are bounded and the learning rates satisfy the Robbins--Monro conditions, the Kushner--Clark theorem ensures that the discrete-time iterates almost surely track the limiting ordinary differential equation (ODE): $\frac{d}{d\xi}\theta^l(\xi)=g(\theta^l(\xi))$, where $\xi$ is the continuous-time variable corresponding to the interpolation of the discrete iteration index $q$.
Let $\epsilon_{\theta^l}(\xi)=\theta^l(\xi)-\theta^{l,\mathrm{NE}}$.
Applying a first-order Taylor expansion of $g(\theta^l(\xi))$ around $\theta^{l,\mathrm{NE}}$ gives
\begin{equation}
\frac{d}{d\xi}\epsilon_{\theta^l}
=
\nabla_{\theta^l}g(\theta^{l,\mathrm{NE}})\,\epsilon_{\theta^l}
+
\Delta(\epsilon_{\theta^l}),
\quad
\|\Delta(\epsilon_{\theta^l})\|
\le
c_2\|\epsilon_{\theta^l}\|^2,
\end{equation}
where $\Delta(\epsilon_{\theta^l})$ contains higher-order terms, and $c_2>0$ is a local constant.
The symmetric part of $\nabla_{\theta^l}g(\theta^{l,\mathrm{NE}})$ is negative definite:
\begin{equation}
\epsilon_{\theta^l}^{\top}
\frac{
\nabla_{\theta^l}g(\theta^{l,\mathrm{NE}})
+
(\nabla_{\theta^l}g(\theta^{l,\mathrm{NE}}))^{\top}
}{2}
\epsilon_{\theta^l}
\le
-\lambda\|\epsilon_{\theta^l}\|^2.
\end{equation}
Then, we construct a Lyapunov function $U(\epsilon_{\theta^l})=\|\epsilon_{\theta^l}\|^2$.
From (31) and (32), its derivative satisfies:
\begin{align}
\frac{d}{d\xi}U
&=
2\,\epsilon_{\theta^l}^{\top}
\nabla_{\theta^l}g(\theta^{l,\mathrm{NE}})\,
\epsilon_{\theta^l}
+
2\,\epsilon_{\theta^l}^{\top}
\Delta(\epsilon_{\theta^l})
\nonumber\\
&\le
-2\lambda\|\epsilon_{\theta^l}\|^2
+
2c_2\|\epsilon_{\theta^l}\|^3.
\end{align}
Then, in a sufficiently small neighborhood of $\theta^{l,\mathrm{NE}}$ where $\lambda-2c_2\|\epsilon_{\theta^l}\|>0$, we obtain $\frac{d}{d\xi}U\le-\lambda\|\epsilon_{\theta^l}\|^2$, implying local exponential stability. Consequently,
\(
\lim_{q\to\infty}\theta_q^l=\theta^{l,\mathrm{NE}},
\quad
\lim_{q\to\infty}\pi_{\theta_q^l}=\pi^{l,\mathrm{NE}}.
\)
\end{proof}

\section{}
\begin{proof}  
Since $\alpha_{u,q}/\alpha_{l,q}\!\to\!0$, the upper-layer DQN evolves on a slower timescale than the lower-layer mean-field PPO. For any fixed $\theta^u$, \textit{Lemma~2} ensures that the lower layer converges to $\theta^{l,\mathrm{NE}}(\theta^u)$, i.e., $
\epsilon_q = \theta^l_q - \theta^{l,\mathrm{NE}}(\theta^u_q) \to 0$. Substituting into the upper-layer update gives
\begin{equation}
    \theta^u_{q+1} = \theta^u_q + \alpha_{u,q} \big( \bar h_u(\theta^u_q) + e_q \big),
\end{equation}
where $\bar{h}_u(\theta^u)=h_u(\theta^{l,\mathrm{NE}}(\theta^u),\theta^u)$ 
and $e_q$ is a martingale noise. 
Following standard stochastic approximation arguments~\cite{borkar2008stochastic}, the upper layer $\theta^u_q$ tracks the ODE $\dot{\theta}^u = \bar h_u(\theta^u)$, which admits a locally stable equilibrium $\theta^{u,*}$. Thus, $\delta_q := \theta^u_q - \theta^{u,*} \to 0$.  

Let $U_l$ and $U_u$ be Lyapunov functions for the lower and upper
layers, respectively. There exist constants
$c_l,c_u,A,B,D_l,D_u>0$ such that
\begin{equation}
\begin{aligned}
U_l(\epsilon_{q+1})-U_l(\epsilon_q)
&\le
-c_l\alpha_{l,q}\|\epsilon_q\|^2
+A\alpha_{u,q}\|\epsilon_q\|^2
+D_l\alpha_{l,q}^2,\\
U_u(\delta_{q+1})-U_u(\delta_q)
&\le
-c_u\alpha_{u,q}\|\delta_q\|^2
+B\alpha_{u,q}\|\epsilon_q\|^2
+D_u\alpha_{u,q}^2,
\end{aligned}
\end{equation}
where $c_l$ and $c_u$ denote the contraction rates of the lower
and upper layers, respectively; $A$ and $B$ characterize the
cross-layer coupling effects; and $D_l$ and $D_u$ bound the
corresponding second-order terms.

Define the composite Lyapunov function as
\(
W_q=U_l(\epsilon_q)+\zeta U_u(\delta_q),
 \zeta>0.
\)
Combining the preceding inequalities yields
\begin{align}
W_{q+1}-W_q
&\le
-\left[
c_l\alpha_{l,q}
-(A+\zeta B)\alpha_{u,q}
\right]\|\epsilon_q\|^2
\nonumber\\
&-\zeta c_u\alpha_{u,q}\|\delta_q\|^2 \quad
+D_l\alpha_{l,q}^2
+\zeta D_u\alpha_{u,q}^2.
\end{align}
Because $\alpha_{u,q}/\alpha_{l,q}\to0$, there exists a sufficiently
large $q_0$ such that, for all $q\ge q_0$,
\(
(A+\zeta B)\alpha_{u,q}
\le
\frac{c_l}{2}\alpha_{l,q}.
\)
Therefore,
\begin{equation}
\begin{aligned}
W_{q+1}-W_q
&\le
-\frac{c_l}{2}\alpha_{l,q}\|\epsilon_q\|^2
-\zeta c_u\alpha_{u,q}\|\delta_q\|^2\\
&\quad
+D_l\alpha_{l,q}^2
+\zeta D_u\alpha_{u,q}^2.
\end{aligned}
\end{equation}

Since the learning rates satisfy the Robbins-Monro conditions, we have 
$\sum_q (\alpha_{l,q}^2+\alpha_{u,q}^2) < \infty$. Thus, the recursion in (37) takes the form $W_{q+1} \le W_q - a_q + b_q$ with $W_q \ge 0$ and $\sum_q b_q < \infty$. By the Siegmund convergence lemma, $W_q$ converges and $\epsilon_q \to 0$ and $\delta_q \to 0$. Therefore,
$(\theta_q^u,\theta_q^l)
\to
\big(\theta^{u,*},
\theta^{l,\mathrm{NE}}(\theta^{u,*})\big)$.
\end{proof}

\bibliographystyle{IEEEtran} 

\bibliography{ref} 









\end{document}